\PassOptionsToPackage{bookmarksnumbered,unicode,colorlinks=true,allcolors=blue,breaklinks=true}{hyperref}
\documentclass[sigconf]{acmart}
\AtBeginDocument{%
  }

\usepackage{colortbl}
\usepackage{soul}
\usepackage{xspace} % for the \sysname
\usepackage[dvipsnames]{xcolor}
\usepackage{xltabular} % For multi-page tables with flexible columns
\usepackage{hyperref}   % For clickable links
\usepackage{booktabs}   % For better line spacing (optional)
\usepackage{longtable}
\usepackage{array}
\usepackage{xurl} % This allows breaks at ANY character

\makeatletter
\g@addto@macro{\UrlBreaks}{\UrlOrds}
\makeatother

\newcolumntype{L}[1]{>{\raggedright\let\newline\\\arraybackslash\hspace{0pt}}p{#1}}
\newcolumntype{R}[1]{>{\raggedleft\let\newline\\\arraybackslash\hspace{0pt}}p{#1}}
\newcolumntype{M}[1]{>{\raggedright\let\newline\\\arraybackslash\hspace{0pt}}m{#1}}

\definecolor{darkgreen}{rgb}{0.0, 0.4, 0.13}
\definecolor{darkgreen2}{HTML}{037171}
\definecolor{lightgreen2}{HTML}{83c5be}
\definecolor{coral2}{HTML}{e29578}
\definecolor{lightlightblue}{rgb}{0.9, 0.95, 1.0}

\definecolor{mustard}{rgb}{0.9, .61, .11}

\copyrightyear{2026}
\acmYear{2026}
\setcopyright{cc}
\setcctype{by}
\acmConference[ASSETS '26]{The 28th International ACM SIGACCESS Conference on Computers and Accessibility}{October 25--28, 2026}{Vila Nova de Gaia, Portugal}
\acmBooktitle{The 28th International ACM SIGACCESS Conference on Computers and Accessibility (ASSETS '26), October 25--28, 2026, Vila Nova de Gaia, Portugal}
\acmDOI{10.1145/3797867.3829024}
\acmISBN{979-8-4007-2521-0/2026/10}

\author[A. Chheda-Kothary]{Arnavi Chheda-Kothary}
\authornote{Work conducted during an internship at the Allen Institute for AI.}
\affiliation{
   \department{Paul G. Allen School of Computer Science \& Engineering}
   \institution{University of Washington}
   \city{Seattle}
   \state{WA}
   \country{USA}
   }
\email{chheda@cs.washington.edu}

\author[L. L. Wang]{Lucy Lu Wang}
\affiliation{
   \department{The Information School}
   \institution{University of Washington}
   \institution{Allen Institute for AI}
   \city{Seattle}
   \state{WA}
   \country{USA}
}
\email{lucylw@uw.edu}

\author[J. C. Chang]{Joseph Chee Chang}
\affiliation{
   \institution{Allen Institute for AI}
   \city{Seattle}
   \state{WA}
   \country{USA}
   }
\email{josephc@allenai.org}

\author[J. Bragg]{Jonathan Bragg}
\affiliation{
   \institution{Allen Institute for AI}
   \city{Seattle}
   \state{WA}
   \country{USA}
   }
\email{jbragg@allenai.org}

\begin{document}

\title[Querying Multimodal Scientific Papers with AI]{Querying Multimodal Scientific Papers with AI: Practices and Preferences Across Blind, Low-Vision, and Sighted Scientists}
% Understanding Querying Practices and Preferences for Multimodal Scientific Documents Across Blind, Low-Vision, and Sighted Scientists

%\author{TBD}

% \renewcommand{\shortauthors}{TBD}

\begin{teaserfigure}
\centering
\includegraphics[width=\textwidth]{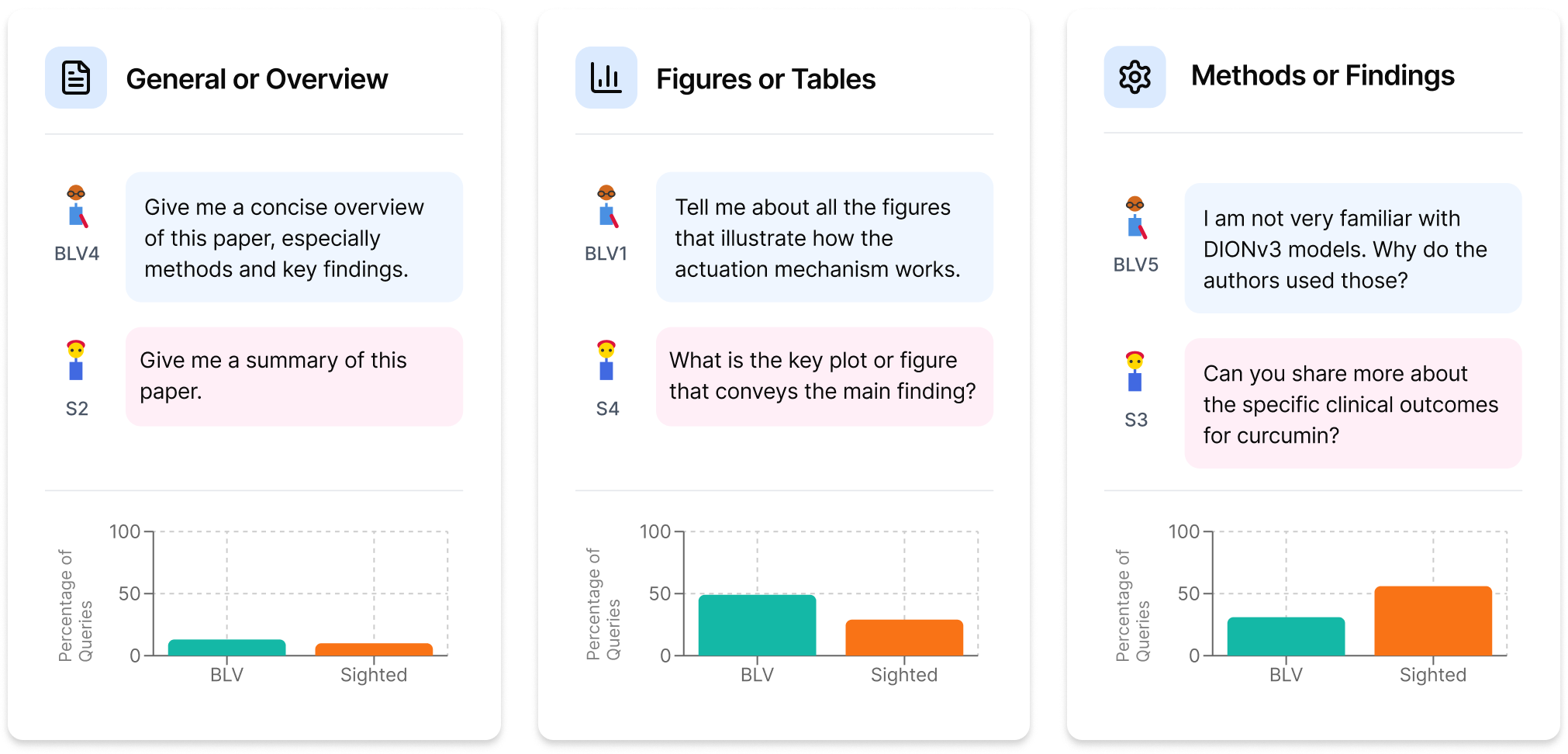}
\Description{Three columns that show the distribution of types of queries participants asked during the study. Column 1 is "General or Overview", and has two chat bubbles, the first from participant BLV4 who asked: "Give me a concise overview of this paper, especially methods and key findings." And the second chat bubble is from participant S2 who asked: "Give me a summary of this paper." Column 2 is "Figures or Tables", and the two chat bubbles here are from BLV1 asking: "Tell me about the figures that illustrate how the actuation mechanism works." and S4 asking: "What is the key plot or figure that conveys the main finding?". Column 3 is "Methods or Findings", the two chat bubbles here are from BLV5 asking: "I am not very familiar with the DIONv3 models. Why do the authors use those?" and from S3 asking: "Can you share more about the specific clinical outcomes for curcumin?". At the bottom, in line with each of the columns, are three bar charts that show the distribution of BLV and Sighted queries for each category. General or Overview has similar distributions around 10\%, Figures or Tables has almost double BLV queries than Sighted queries (roughly 50\% to 29\%), and Methods or Findings has almost double Sighted queries than BLV queries (56\% to 31\%).}
\caption{We found three main categories of queries participants asked using QA systems---(1) General or Overview, (2) Figures or Tables, and (3) Methods or Findings. Within these, BLV participants asked almost double the percentage of Figures or Tables queries as sighted participants. Contrastingly, sighted participants asked almost double the percentage of Methods or Findings queries as BLV participants. Both groups had a similar strategy of first situating themselves in the overall structure and goal of a paper with General or Overview queries.}
\label{teaser}
\end{teaserfigure}

\begin{abstract}
Visual diagrams, figures, and tables are central to scientific papers, and convey information beyond what is captured in text. While blind or low-vision (BLV) scientists have traditionally relied on static alternative text to access figures in papers, the rise of artificial intelligence (AI) has made interactive question-answering (QA) a feasible paradigm for visual exploration; yet little is known about how scientists use visual QA in practice or how to improve its accessibility. In this work, we interview five BLV and five sighted scientists across different STEM fields to understand how they use two AI tools, \textit{ChatGPT} and \textit{Gemini}, to query multimodal scientific documents. Our findings characterize how scientists review multimodal content, including existing practices (along with accessibility workarounds) for engaging with visuals, and feedback on the suitability of AI-generated responses to multimodal queries. We further find that vague or incomplete image descriptions, as well as incorrect AI outputs more broadly, can cause both BLV and sighted scientists to abandon AI workflows. To support future research, we additionally contribute a dataset of 115 queries and responses from our participants' interactions with the AI tools for papers in their field. We close by discussing implications for AI-powered scientific QA systems, emphasizing considerations for access across abilities and domains.
% \jb{unclear on why access across abilities is an important goal - needed to demonstrate necessary to interview sighted scientists}.
\end{abstract}





\maketitle

\section{Introduction}
\begin{quote}
    \textit{``Scientific communication is very visual. You are not going to publish a paper if you have bad graphs; it doesn't matter how good your results are. It is much easier for you to just look at something and get the information than to scroll down to a very heavy explanation in the paper. So the nature of science itself is inherently non-accessible.''}---\textbf{BLV5}, a blind scientist
\end{quote}
% State of the world
Scientific papers are inherently multimodal, combining text with visual elements such as diagrams, graphs, images, and tables to communicate complex ideas. These visual components are not merely supplementary; they encode relationships and abstractions that are difficult to convey through text alone. Prior work across domains including medicine and computer science has highlighted the critical role of figures and visualizations in supporting comprehension and knowledge transfer \cite{Jeyaraman2024DecodingResearchWithAGlance, Evagorou2015TheRoleOfVisualRepresentations, Millar2022TheRoleOfVisualAbstracts}. Notably, \citet{Lee2016ViziometricsAnalyzingVisualInfo}, in a large-scale analysis of over 8 million PubMed figures, found significant correlation between the use of visual information and scientific impact, underscoring the central role of multimodal communication in shaping how research is understood and disseminated.

% The big BUT
For scientists who are blind or low-vision (BLV), engaging with scientific multimodal content such as figures and tables can be challenging. Historically, BLV individuals have relied on alternative text (alt-text), a succinct description of an image usually written by a paper's authors, to interpret visuals. But quality alt-text is challenging to write \cite{Mack_QualityAltText}, and alt-text in scientific papers remains scarce, as shown by \citet{KumarAndWang} who found that only 8.5\% of 20,000 papers across 19 fields included any alt-text, and many of those were in fact insufficient. Even when alt-text provides some level of description, it is inherently static and can be inadequate for scientific use, where figures are more complex and the goals of figure understanding may shift depending on the specific scientific tasks and domains of the readers.

Prior work exploring complex image representation has therefore called for moving beyond one-size-fits-all image descriptions \cite{Stangl_OneSizeFitsAllImageDescriptions} toward richer, more flexible representations of visual content, such as interactive Question-Answering (QA) \cite{Morris_RichRepresentations}. Flexible representations of visual content can benefit sighted individuals in addition to BLV individuals, as \citet{Li_GeoVisA11y} highlight with geovisualizations---in their work, sighted participants preferred a BLV-accessible QA system when visualizations became more complex. However, description and interaction preferences are not uniform across groups; \citet{LundgardSatyanarayanAccessibleViz} found that blind and sighted readers differed significantly in which kinds of semantic content they considered most useful. While the rapid rise of artificial intelligence (AI) has made visual QA increasingly accessible, we still do not know how well these systems support scientists with diverse visual access needs in engaging with scientific documents, where visual content is complex, domain-specific, and intertwined with text. Addressing this gap is key to designing inclusive scientific QA systems.

% Therefore we did
To characterize how current AI-powered QA systems support (or fail to support) access to multimodal scientific content in practice, we investigate the question: \textit{how do scientists with diverse visual access needs query multimodal scientific papers using AI systems, and what practices and preferences shape this interaction?} We use the term \textit{multimodal} in the present work to describe interwoven text, images, and tables in papers. To explore our research question, we conducted semi-structured interviews with 10 scientists across various STEM domains, including medicine, neuroscience, computer science, and more, to learn about how they use AI systems to query over scientific documents. Five of our participants identified as BLV while the other five identified as sighted. Prior work has highlighted the limitations of siloed, population-specific accessibility technologies and their poor integration into broader real-world systems \cite{Mankoff_DisabilityCriticalInquiry, Wobbrock_AbilityBasedDesign}, with recent work showing similar fragmentation in AI accessibility efforts \cite{Moharana_2025}. We therefore examine how multimodal QA systems can support both blind and sighted scientists, capturing a broader range of querying strategies. The interviews happened over Google Meet, and consisted of: (1) an introductory discussion to learn about current workflows with reading papers, (2) the use of two popular AI tools, \textit{ChatGPT} and \textit{Gemini}, for participants to query papers from their specific domains, and (3) a closing discussion reflecting holistically on the role of AI-powered QA systems for science. We collected and analyzed participants’ queries and responses alongside the broader interview data.

% We include both BLV and sighted scientists to examine similarities and differences in querying practices. Our goal is not to treat sighted use as a baseline \cite{Mack_WhatDoWeMeanByAccessibilityResearch}, but to understand the range of strategies that emerge across diverse visual access needs. With this approach, we aim to enable designing AI systems that are not overfit to a single population, but instead support more inclusive and flexible multimodal interactions. 

% The key findings are
%\l{maybe a sentence or two here summarizing the three stages of the interview study and then through analysis, our findings show...} 
Our findings show that while multimodal content is central to scientific understanding, current workflows remain highly inaccessible for BLV scientists. For example, BLV scientists reported specialized workarounds, which include extracting text from papers with custom scripts, seeking figure descriptions from sighted colleagues, and leveraging AI-powered QA systems. We also categorize and present the queries that BLV and sighted participants asked of papers in their domain using ChatGPT and Gemini. BLV scientists used these systems as a primary bridge to foundational visual content (49\% of their queries were about figures or tables), while sighted scientists leveraged them more to accelerate synthesis around specific methods or findings (56\% of queries), using the agents to extract specific technical parameters and clarify complex intuitions behind methods. Despite these different usage patterns, we observed a shared funneling strategy where all researchers moved from initial broad queries to granular deep-dives. However, across both groups, we identified a critical gap: AI-generated descriptions of multimodal content are often vague or inconsistent, failing to meet the precision required for scientific work. While this led to loss of trust in AI systems for both BLV and sighted scientists, we found that this lack of precision had a greater impact on BLV scientists, as they could not quickly verify inaccuracies when the paper itself was inaccessible, unlike their sighted counterparts.

% The contributions of this work are
In sum, we contribute: (1) empirical findings from interviews with five BLV and five sighted scientists using QA to engage with multimodal scientific documents, and (2) a dataset of 115 queries and responses from the interviews to support understanding and development of AI-powered QA systems. 

\section{Related Work}
We present prior work in supporting BLV scientists for paper reading, rich representations of complex visuals that introduce the QA paradigm, and work in benchmarking scientific document QA. 

\subsection{Technology Supporting BLV Scientists}
Research on supporting BLV scientists first emphasizes the persistent inaccessibility of scholarly PDFs \cite{Mowar_iTagPDF, Wang_SciA11y, KumarAndWang}. \citet{Wang_ImprovingA11yofSciDocs} reveal that only 2.4\% of 11,397 scientific PDFs satisfy all of their accessibility criteria, and introduce SciA11y \cite{Wang_SciA11y}, which converts scientific PDFs into accessible HTML and adds navigational features for screen reader users. More recently, \citet{KumarAndWang} characterize what they call a \textit{``new accessibility crisis''} in scholarly PDFs, finding that less than 3.2\% of 20K papers satisfy all accessibility criteria they measured and that scientific paper accessibility has been declining after 2019. Beyond PDF accessibility, navigation of papers itself remains a major challenge for BLV readers: \citet{Park_TeamSourcedHyperlinks} find that low-vision readers want navigation features they could invoke throughout a paper that preserved the paper’s original narrative while helping unpack content.

Prior work on nonvisual STEM access for scientists also includes multimodal graph systems that combine haptics, audio, and structured feedback to support graph interpretation \cite{Yu_EvaluationMultimodalGraphs, Walker_UniversalDesignAuditoryGraphs}. More recent work has also established design guidelines for touchscreen-based multimodal graphics, further demonstrating the value of dynamic, nonvisual representations for complex STEM content \cite{Gorlewicz_DesignGuidelinesTouchGraphics}. We position this literature as complementary to our work: rather than broadly studying nonvisual STEM figure access, we focus specifically on how scientists use multimodal QA systems to interpret figures and other visual content within scientific papers.

More relevant to the present work is research on making scientific visuals accessible through alt-text and better image descriptions. \citet{Williams_QualityAltTextInComputing} examine how to support authors in creating higher-quality alt-text for computing publications, while \citet{Chintalapati_AltTextsDataset} share a dataset of alt-texts from human-computer interaction (HCI) publications to characterize their semantic content and support more descriptive scientific figure descriptions. \citet{Singh_FigurA11y} extend this line of work with \textit{FigurA11y}, an interactive system that generates scientific alt-text and suggests revisions for paper authors to improve the alt-text of their figures. More recently, \citet{Iwamoto_GenAIImageDescriptionsForScienceFigs} highlight how university disability service professionals write descriptions for HCI science figures without subject expertise, sharing the difficulty of producing useful scientific descriptions when domain knowledge is limited. They also explore if and how generative AI can support disability professionals to improve scientific alt-text. While this prior work has focused on traditional alt-text and image descriptions, in the present work, we shift toward QA for more interactive multimodal understanding.

\subsection{Beyond Alt-Text: Rich Representations of Complex Visuals for BLV Individuals}
Prior work argues that static alt-text is usually insufficient for supporting rich engagement with complex visuals. \citet{Morris_RichRepresentations} call for supplementing alt-text with richer digital representations such as QA, letting BLV screen reader users navigate visual content at multiple levels of detail. In visualization, \citet{LundgardSatyanarayanAccessibleViz} highlight strategies for effective natural language descriptions of complex visuals, and they find that blind and sighted readers prefer different descriptions. Other work explores tactile and hybrid representations---recent work on tactile charts shows that BLV participants are still predominantly limited to alt-text for complex visualizations, and \citet{He_TactileCharts} propose tactile chart designs as a richer pathway for comprehension and learning. \citet{Tsutsui_TouchAndTalkScienceCommunication} examine how interactive dialogue can be paired with tactile exploration in science communication.

More recent interactive digital systems for BLV visual access have moved further toward QA. \textit{VoxLens} \cite{Sharif_VoxLens} is a browser plugin that lets users ask natural-language questions of online charts and receive spoken answers, and \citet{Kim_ChartQA} conduct a Wizard-of-Oz study on chart QA for BLV participants, collecting user queries and mapping them to taxonomies. In the present work, we similarly collect user queries specific to multimodal scientific papers, which we categorize and present. Most recently, \textit{GeoVisA11y} \cite{Li_GeoVisA11y} show that QA supported by large language models (LLMs) can improve reading, analysis, interpretation, and navigation of geovisualizations, and that BLV and sighted participants exhibit distinct queries while still identifying similar patterns from the visualization. 

Commercial tools such as \textit{Be My AI} \cite{be_my_ai_2024} and \textit{Seeing AI} \cite{seeing_ai} have also popularized conversational image understanding for everyday tasks and provide the opportunity for BLV users to follow up through QA. Recent research by  \citet{Penuela2026HowMLLMsSupportVisualInfoAccess} also explores multimodal LLMs for visual assistance, highlighting that BLV users prefer to ask multiple questions about a particular image even after an initial description and want AI to be aware of their unique goals. In the present work, we extend these explorations of richer multimodal representations leveraging QA and conversational AI to querying scientific papers. Furthermore, by involving both BLV and sighted scientists (taking a similar approach as \citet{Li_GeoVisA11y}), we are able to study how QA systems support multimodal scientific sensemaking across diverse visual access needs, where figures and tables are interpreted within the context of the full paper rather than as standalone charts and visualizations.

\subsection{Benchmarks and Datasets for Multimodal Scientific QA Interactions}
Beyond the HCI research investigating QA for complex visualizations, such as VoxLens and GeoVisA11y, other work exploring multimodal QA has primarily investigated the space through benchmarking model performance. \textit{SPIQA} by \citet{NEURIPS2024_SPIQA} introduces a large-scale dataset for answering questions about figures and tables in scientific papers, with 270,000 synthetically-generated questions over plots, charts, tables, and schematic diagrams. \textit{MISS-QA} \cite{Zhao2025CanMFSchematicDiagrams} similarly studies information-seeking QA over schematic diagrams in scientific papers, while \textit{M3SciQA} \cite{Li2024M3SciQAAM} extends multimodal QA to multi-document scientific reasoning. \textit{MathVerse} \cite{Zhang2024MathVerseDY} evaluates whether multimodal models actually use diagram information in visual math problems rather than relying on text-based clues for math understanding. Across these efforts, the focus is on how well models answer questions about multimodal documents that combine text, figures, and math; while important, these efforts typically use synthetic data and do not capture how people with diverse visual access needs want to use such systems in practice. 

Even the limited BLV-aligned work in this area remains centered on dataset construction rather than interactive use. \citet{Kang2025SightationCL} present a BLV-aligned dataset spanning 5,000 diagrams and 137,000 samples of diagram descriptions called \textit{Sightation} by having sighted annotators assess descriptions generated by vision-language models, rather than collecting real-world QA interactions from BLV users. Our work complements these resources by examining multimodal scientific QA as an accessibility interaction paradigm, as we collect real user data and feedback from BLV and sighted scientists on QA interactions over scientific papers.

\section{Study Methods}
To evaluate benefits and limitations of existing QA systems, and to understand querying strategies across BLV and sighted scientists, we conducted semi-structured interviews with 10 STEM researchers, which we report on below. This study was reviewed by the Institutional Review Board (IRB) at our institution and deemed exempt under applicable federal regulations.

\subsection{Participants}
We recruited 10 scientists (Table \ref{Participant_Table}) across various STEM fields using email lists. Five participants were BLV and five were sighted; all were 18 years of age or older, and all engaged with scientific papers at least a few times per week. Participants were located across the United States, so interviews took place over Google Meet. 
%\jb{Was there a diversity of amount of experience/seniority/role?}

\begin{table*}[t]
\centering
\LARGE % Increase the font size
\setlength{\extrarowheight}{5pt} 
\resizebox{\textwidth}{!}{ % Resize the table to fit the width of the page
% Define row colors here
\rowcolors{2}{white}{gray!10}
\begin{tabular}{cccllll}
\toprule 
\rowcolor{white}  % header row color if needed
\textbf{PID} & \textbf{Age Range} & \textbf{Gender} & \textbf{Degree of Vision Loss} & \textbf{Field of Research} & \textbf{Role} & \textbf{Highest Degree Completed} \\
\midrule

BLV1 & 18-24 & M & Totally Blind/No Usable Vision & Computer Science & PhD Student & Bachelor's Degree \\

BLV2 & 25-34 & F & Legally Blind & Science Education & PhD Student & Master's Degree \\

BLV3 & 35-44 & F & Legally Blind & Marine Geology & PhD Student & Master's Degree \\

BLV4 & 25-34 & M & Some Vision Loss & Computer Science & PhD Student & Bachelor's Degree \\

BLV5 & 25-34 & M & Totally Blind/No Usable Vision & Neuroscience and Computer Science & PhD Student & Master's Degree \\

S1 & 25-34 & M & Sighted & Materials Science & Industry Researcher & Doctoral Degree \\

S2 & 35-44 & M & Sighted & Medicine (Pediatric Infectious Diseases) & Clinical Researcher & Doctoral Degree \\

S3 & 35-44 & F & Sighted & Integrative Medicine (Cancer Therapy) & Program Manager & Master's Degree \\

S4 & 25-34 & F & Sighted & Neuroscience & Industry Researcher & Doctoral Degree \\

S5 & 25-34 & F & Sighted & Computer Science & PhD Student & Master's Degree \\
\bottomrule
\end{tabular}
}
\caption{Self-reported demographics of study participants, including information about their vision loss, fields of research, and role. BLV1-BLV5 are the BLV participants, S1-S5 are the sighted participants.}
\label{Participant_Table}
\vspace{-0.8cm}
\end{table*}

\subsection{Procedure}
Participants filled out an online pre-study questionnaire prior to the session to answer questions about demographics, field of research, frequency of engaging with papers, and use of any AI or accessibility tools for parsing papers. Each session consisted of three parts: (1) an opening semi-structured interview, (2) using two AI tools, ChatGPT and Gemini, to query against one unfamiliar paper and one familiar paper from participants' domains, and (3) a closing discussion. Sessions lasted at most 90 minutes over a Google Meet call, and participants were compensated \$150 for their time. 

During the opening interview, we asked participants about their role, research responsibilities, and how they prefer to engage with multimodal papers in their field (including tools they use, paper reading practices, and when/how they engage with figures or tables). Next, participants used the free versions of ChatGPT and Gemini (at the time of the studies, the backing models for these were GPT-5 and Gemini-2.5 respectively) to query two scientific papers on their own computers. We chose these two AI tools after comparing several options---ChatGPT \cite{chatgpt_overview_2026}, Gemini \cite{gemini_about_2026}, Claude \cite{claude_overview_2026}, Ask AlphaXiv \cite{alphaxiv_2026}, Qwen Chat \cite{qwen_chat_2026}, and Poe \cite{poe_2026}---based on their accessibility, functionality support in the free tier, and the ability to upload large PDF documents to the tool. We asked participants to share their screen as they used each of the tools, and we also asked BLV participants using screen readers to share system audio. 

Prior to the session, participants downloaded PDFs of one paper from their field that they had previously read (familiar) and one they had not (unfamiliar). Because participants brought papers from their own fields, we did not standardize whether the PDFs contained alt-text. As a result, only three of the 10 papers from BLV participants had alt-text. In practice, BLV participants rarely engaged with alt-text during the sessions beyond a single instance where BLV1 queried for all figures in a paper and asked the QA system for associated alt-text from the PDF (Appendix \ref{115_Queries}). 

Participants completed four conditions: each tool paired with each paper, with the order counterbalanced across participants. Three participants completed only three of the four conditions due to time constraints, but all still tested both a familiar and an unfamiliar paper. Participants could ask the AI any questions they wanted about the uploaded paper. We did not provide a predefined set of queries, aside from asking that they make at least one query about a figure or table so we could gather targeted feedback on multimodal content descriptions. All queries were typed, and BLV participants used their screen readers and usual text-entry workflows for inputting queries. 

When using the first tool with a given paper, participants could query the uploaded PDF but were asked not to open the paper in a separate PDF viewer on their own machine (\textit{e.g.}, Adobe Acrobat or Preview), meaning they could not initially validate responses against the source document outside the AI tool. When they transitioned to the second tool for that same paper, they were allowed to open the PDF in any viewer they preferred for checking or validation, which allowed the research team to compare querying behaviors with and without direct access to the source material. We did not impose strict time limits for switching tools; participants were encouraged to move on after roughly 10–15 minutes, but could continue if they wished, particularly so as not to constrain BLV participants who sometimes needed more time. We asked participants to use the think-aloud method \cite{ericsson1993protocol} to describe their choice of queries and feedback to AI responses as they worked with the different papers and tools. Finally, we asked participants to reflect holistically on using the two AI tools to query papers.

We included variations in tools, paper familiarity, and PDF-access in our study to broaden the range of interaction contexts participants encountered, rather than to support condition-level comparisons. Accordingly, our analysis does not make comparative claims across tool, familiarity, or PDF-access conditions, but instead uses them to surface a wider range of querying practices.

\subsection{Analysis}
The first author reviewed the 10 user study transcripts, and arrived at an initial set of 17 codes using inductive and deductive coding \cite{Braune_And_Clarke}. These codes were reviewed by the research team, and then presented along with two user study transcripts (one sighted participant, S1, and one BLV participant, BLV2) to an external researcher, who reviewed them and the first author's comments and rationales, resulting in one additional code. The first author and external researcher then engaged in peer debriefing \cite{PeerDebriefing}, and condensed the 18 codes into eight themes which were then grouped under three higher-level themes following thematic analysis practices \cite{Braune_And_Clarke}. We additionally used peer debriefing along with affinity diagramming \cite{beyer1998contextual} within the four members of the research team to categorize the queries from participants engaging with the AI tools and extract relevant insights, presented as a part of the themes. Our analysis method follows from similar past qualitative HCI studies and best practices \cite{Chheda_ASSETS, Li_SidewalksBigDeal, VizXpress_Zhang, McDonald_NormsAndGuidelines}.

\section{Findings}
We present our findings organized under three higher-level themes: (1) current research workflows which highlight how BLV and sighted scientists engage with multimodal papers today, (2) how participants in our study used multimodal QA (through ChatGPT and Gemini) for querying against papers in their domain, and (3) feedback around validation, trust, and privacy when using AI-powered QA systems for scientific tasks.

\subsection{Current Research Workflows and Engaging with Multimodal Content}
We share how BLV and sighted scientists currently undertake paper finding and reading, including their preferences around engaging with figures and tables in papers, and accessibility workarounds that BLV scientists employ.

\textbf{Holistic practices of working with papers.}
Our participants take several approaches to finding papers, including using tools like Google Scholar\footnote{https://scholar.google.com/}, Semantic Scholar\footnote{https://www.semanticscholar.org/}, or PubMed\footnote{https://pubmed.ncbi.nlm.nih.gov/}, using university catalogs and resources including help from librarians, and parsing conference-specific websites for proceedings. All participants reported using some form of a PDF reader (Adobe Acrobat, MacOS Preview, Safari Reader) for reading papers. Four out of five of the BLV participants are regular screen reader users---BLV1, BLV2, BLV3, BLV5---and they use their screen readers along with PDF reader software to access papers. The fifth BLV participant, BLV4, uses a screen reader for paper parsing when he is tired, but primarily relies on OS magnification tools.

Seven participants have attempted to use AI tools (ChatGPT, Gemini, Claude, NotebookLM, JAWS PictureSmart) with six of those participants saying they have incorporated AI usage into their regular research workflows; S3, who has tried several AI tools in the past but was unhappy with her experience stated: \textit{``When you’re doing systematic reviews, you need to be comprehensive; [AI tools] are not comprehensive.''} The six scientists who do regularly use AI tools for paper reviews have different approaches, including: paper summarization (BLV1, BLV4), figure interpretation via QA beyond alt-text (BLV1, BLV2), understanding specific results, jargon, or arguments (BLV1, BLV2, BLV5, S5), and domain-specific needs such as finding literature to inform a clinical decision (S2). BLV1 and BLV5 also described taking screenshots of specific PDF pages or figures from a paper and feeding those to AI systems such as ChatGPT for more targeted QA interactions. 

\textbf{When and why scientists engage with multimodal content.}
Both sighted and BLV participants emphasized the importance of figures and tables in scientific papers across their various domains. Sighted participants (S1, S2, S4) explained that they often look at figures first to \textit{``walk through''} new methods or quickly grasp the main hypothesis or findings. S4 explained how in her field, neuroscience, the visuals are often seen as the most important part of the paper, and text serves as a secondary confirmation. S1, a materials science researcher, described how figures are essential for deciphering complex data, and he will often use WebPlotDigitizer\footnote{https://automeris.io/} to manually tabulate data from graphs to understand the data powering the visuals. While in some domains such as materials science, a traditional system diagram may be seen as unnecessary (S1), others such as S5, a computer scientist, described exploring overall figures and system architecture diagrams as a large part of reading a new paper. Researchers from other domains, such as medicine (S2, S3) choose to focus more on tables rather than figures.

For BLV researchers, the decision to engage with a figure often depends on its potential to provide specific details that text cannot easily convey (BLV1, BLV3). BLV1, a computer scientist, described frequently skipping the introductory figure if the abstract already covers the high-level summary. BLV2 and BLV3 prioritize understanding complex charts, such as map-based seismic graphs or 3D chemistry charts, where spatial relationships conveyed in those graphics are critical to understanding the paper results. This engagement is driven by a need for precision that general descriptions, captions, and existing alt-text often lack; for instance, BLV3, a marine geology researcher, needs to understand exactly how a data point relates to a specific geographic region. BLV2, who works in chemistry and biology education, has to have a precise understanding of graphs and trends.  

Participants noted that even when using AI for summaries, the systems often fail to properly interpret the figures beyond what is described in text, forcing BLV scientists to seek out specialized, expert-level descriptions to ensure they aren't missing the \textit{``salient points''} necessary for scientific standards (BLV2, BLV3, BLV4). As one example, BLV2 described a previous QA interaction she had with an AI tool for graph interpretation where an AI response inaccurately presented a granular detail she needed: 
\begin{quote}
    \textit{``That more interactive [QA] component can be helpful, but [AI] second-guesses itself a lot. I've had ones where I have measurements on the screen and you're like, `Okay, that graph, where does it peak?' And the [AI] is like, `About 10 degrees.' And I'm like, `Are you sure? That doesn't sound right.' And it's like, `Oh, you're right. It's actually 15 degrees.' Then you're like, `Okay, that sounds a little closer.' And they're like, `Looking back, I believe it's 10 degrees.' ''}---\textbf{BLV2}
\end{quote}

To get more detailed and relevant descriptions, BLV1 and BLV5 frequently turn to their sighted colleagues and labmates. BLV3 asks her sighted husband (also a scientist) for help describing figures: 
\begin{quote}
    \textit{``I ask him what is this figure and then, he has a science background too, so he can give me the scientific valid points; versus if I go to like my disability office and they have a history degree or a psychology degree, they're not going to understand what are the scientific values that I need.''}---\textbf{BLV3}
\end{quote}

\textbf{Existing accessibility workarounds and prompting strategies.}
Although PDFs remain the dominant format for scientific communication, participants described their continued inaccessibility. As a result, BLV scientists have developed a range of workarounds and prompting strategies. BLV1 has a script for text extraction: \textit{``I wrote a Python script just to extract the text from the PDF''}. BLV5 has used prompts to reconstruct visual information, \textit{i.e.}, asking systems to explain how a figure was generated to get mathematical or code-based representations: \textit{``I love having math representation of things even more than code''}. For complex visuals, BLV2 and BLV5 use an embosser to create tactile renditions of images; but this can be time consuming, with BLV2 saying it can take \textit{``a minimum of two to three hours per picture.''} Participants also had prompting strategies to mitigate unreliable outputs, including explicitly discouraging hallucinations (BLV3) and asking the AI system to cross-check responses against external internet sources (BLV2).

\subsection{Multimodal QA in Practice}\label{MMQA}
We examined how participants used QA systems in practice with papers from their domain using ChatGPT and Gemini, focusing on the types of queries they posed and how they evaluated system responses. The insights presented below reveal both the breadth of interaction strategies and the limitations of current AI-generated descriptions. The full set of 115 queries and AI responses can be found in the Appendix (Section \ref{115_Queries}) and supplementary materials.

\textbf{Types of queries and information needs.}
To examine how scientists with different visual access needs and disciplinary backgrounds use QA systems for paper comprehension, we categorized the 115 queries into four types: (1) General or Overview, (2) Figures or Tables, (3) Methods or Findings, and (4) Miscellaneous (encompassing queries around future work, author backgrounds, and ad-hoc corrections). BLV and sighted participants all followed a consistent funnel approach, starting with general or overview queries (\textit{e.g.,} \textit{``Summarize the paper''} or \textit{``Describe the key findings''}) to situate themselves before pivoting to queries around specific figures and methods. After starting with an overview, there were then clear differences in query intent between BLV and sighted participants. The five BLV participants primarily used QA systems to access visual content, with 49\% of their queries focused on figures and tables. These queries often asked for descriptions of chart axes, trends, data relationships, or system specifics. In contrast, sighted participants directed only 29\% of their queries toward visual elements, instead focusing on synthesizing methods or locating specific findings (56\%). Notably, four out of five BLV participants (BLV1, BLV3-5) had at least one query where they asked for descriptions of \textit{all} figures in the paper to get an overview of the available multimodal content to dive into further.

Domain-specific needs further shaped these behaviors. Participants in materials science and medical research primarily issued queries to extract experimental details and outcomes. For example, S1 (materials science) treated a research paper as a technical manual, asking, \textit{``my shaft diameter is 3" with differential pressure of 8 psi. What would my leak rate be?''} and querying the specific mathematical relationship between leak rate and pressure. Participants in computer science and neuroscience queried for more structural and conceptual clarification. For instance, S4 (neuroscience) sought \textit{``intuitions''} for complex methods like \textit{``avalanche analysis,''} while BLV4 (CS) requested a \textit{``timeline-style summary''} to understand how pilot study results \textit{``refined the main study protocol.''} The strongest reliance on visual interpretation emerged from BLV3 who works in marine geology, a more spatially oriented domain---she dedicated 67\% of her queries to figure descriptions.

\begin{figure*}[h]
\centering
\includegraphics[width=\textwidth]{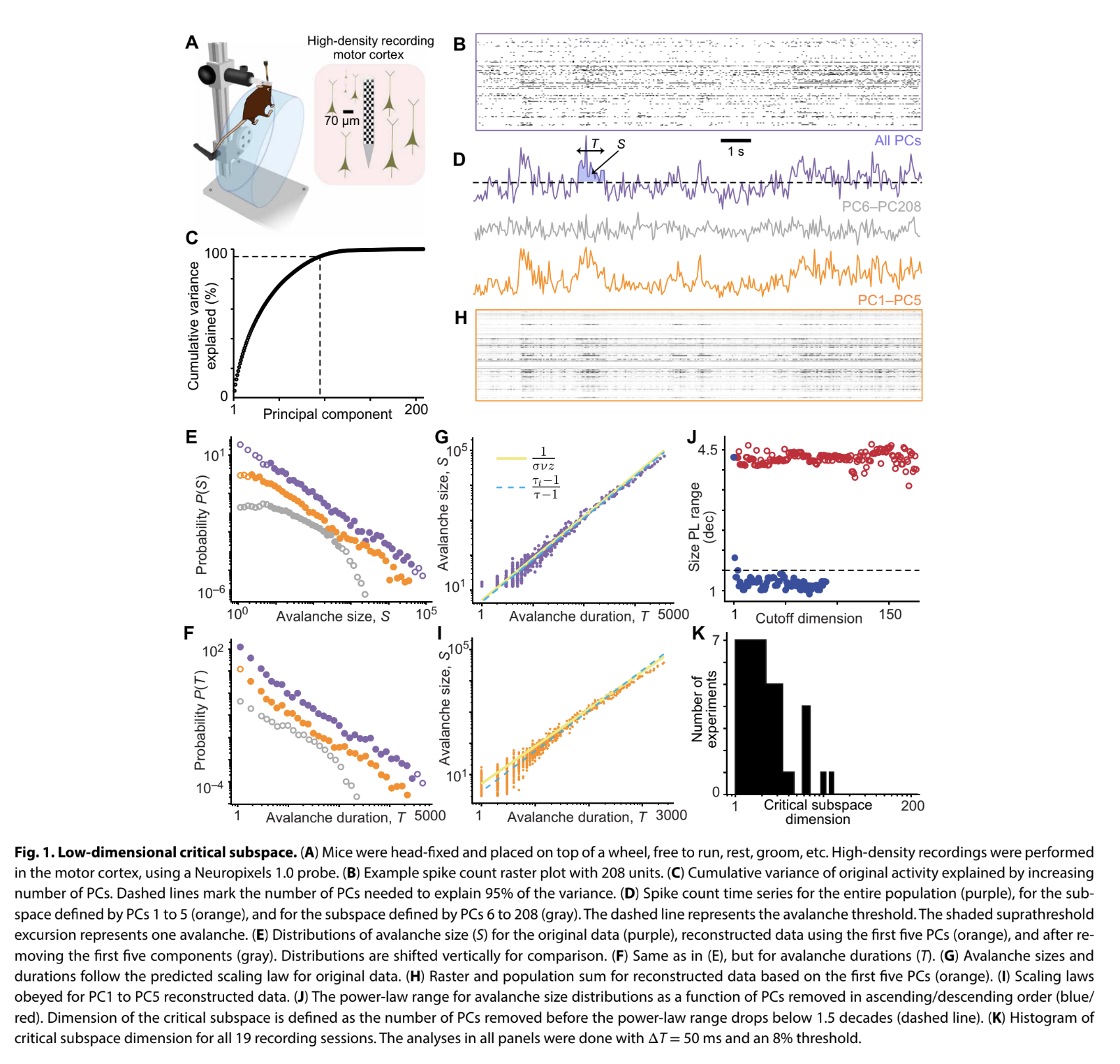}
\Description{A figure from a neuroscience paper that shows the Low-dimension critical subspace. There are 11 sub-figures, labeled A-K. A has an image of a mouse doing an activity, and B-K are different chart representations of the activity. The original figure caption is at the bottom.}
\caption{An example of a complex, multi-part figure from a neuroscience paper, ``Low-dimensional criticality embedded in high-dimensional awake brain dynamics'' by \citet{fontenele2024low}. S4 used QA to understand subpart J of this figure, asking: ``Could you interpret Fig 1J for me?''}
\label{S4_MultiPartFigure}
\end{figure*}

We also observed that sighted participants often used multimodal QA to verify consistency between textual descriptions and visual content, compared to BLV queries which were more oriented towards figure understanding. For example, S5 asked about \textit{``specific differences between the methods as described in the text and the Figure 1b diagram''} to identify mismatches between text and figures. Similarly, S4 used the system to interpret complex sub-figures (\textit{e.g.}, Fig 1J as shown in Figure \ref{S4_MultiPartFigure}) and confirm whether reported trends aligned with her own interpretation of the plots. However, the latter verification strategy used by S4 relied on access to the paper, and was limited during the portion of the study where she used the QA system without referencing the source document. We elaborate more on validation across the agent and the original paper as well as ways in which trust is lost in the QA process in Section \ref{ValidationTrustPrivacy}.

\begin{figure*}[h]
\centering
\includegraphics[width=\textwidth]{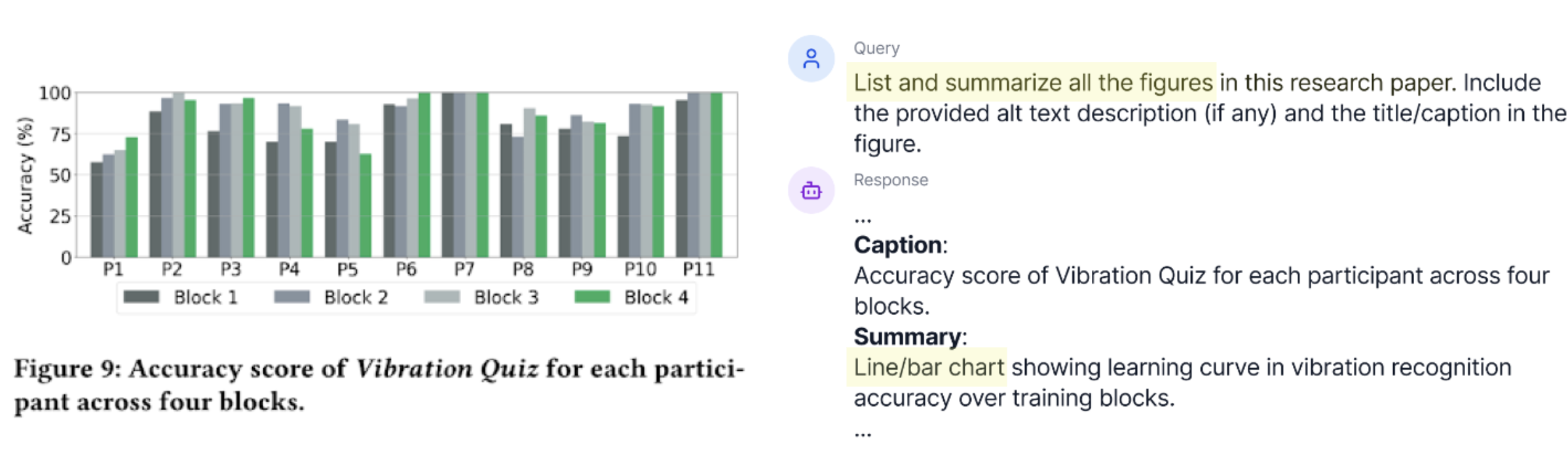}
\Description{Figure 9 from the Typing Haptically paper is shown on the left, which is a bar chart representation of accuracy score across participants and blocks. The original caption is below the figure. On the right is the user query asking to list and summarize all the figures in this paper, below which is the model response that includes the figure caption and the summary, saying: "Line/bar chart showing learning curve in vibration recognition accuracy over training blocks." Line/bar chart is highlighted, showing the source of user confusion in the AI response.}
\caption{Ambiguity in AI descriptions of charts: the description for Figure 9 from the paper ``Typing Haptically: Towards Enabling Non-auditory Smartphone Text Entry with Haptic Feedback for Blind and Low Vision Users'' by \citet{TypingHaptically} led to confusion for BLV1 because the chart is described as a line/bar chart.}
\label{BLV1_LineBarChart}
\end{figure*}

\textbf{Optimal multimodal descriptions, and where current AI responses fall short.}
All participants reported that AI-generated responses to their queries of multimodal scientific papers were frequently too vague to support their academic work. Several BLV participants highlighted the ambiguity in chart descriptions. For example, BLV1 stated after reading a chart description in a QA response, that he was \textit{``unsure if it's a line chart or a bar chart''} (Figure \ref{BLV1_LineBarChart}), while BLV3 emphasized the inability to analyze any trends due to the AI response omitting ranges and scales on axes. BLV4 noted that lacking context early in the AI response made details harder to parse, emphasizing the need for a brief overview of figures at the start of descriptions (\textit{e.g.}, \textit{``two side-by-side plots''}). BLV3 criticized the use of relative terms like \textit{``high''} to describe trends without a baseline for comparison. BLV and sighted participants holistically lamented that current AI models often fail to reach the \textit{``nuance''} (S3) required for scientific inquiry, falling back on surface-level text and visual semantic recognition rather than deep understanding and interpretation of the figure. Participants also expressed a desire for figure and table descriptions that provide new insights beyond what is in the text or captions. As S5 stated: \textit{``I [want to] ask it to pick out trends, then direct my attention [to] trends that I wouldn't have noticed... I could see that being super useful.''}

\begin{figure*}[h]
\centering
\includegraphics[width=\textwidth]{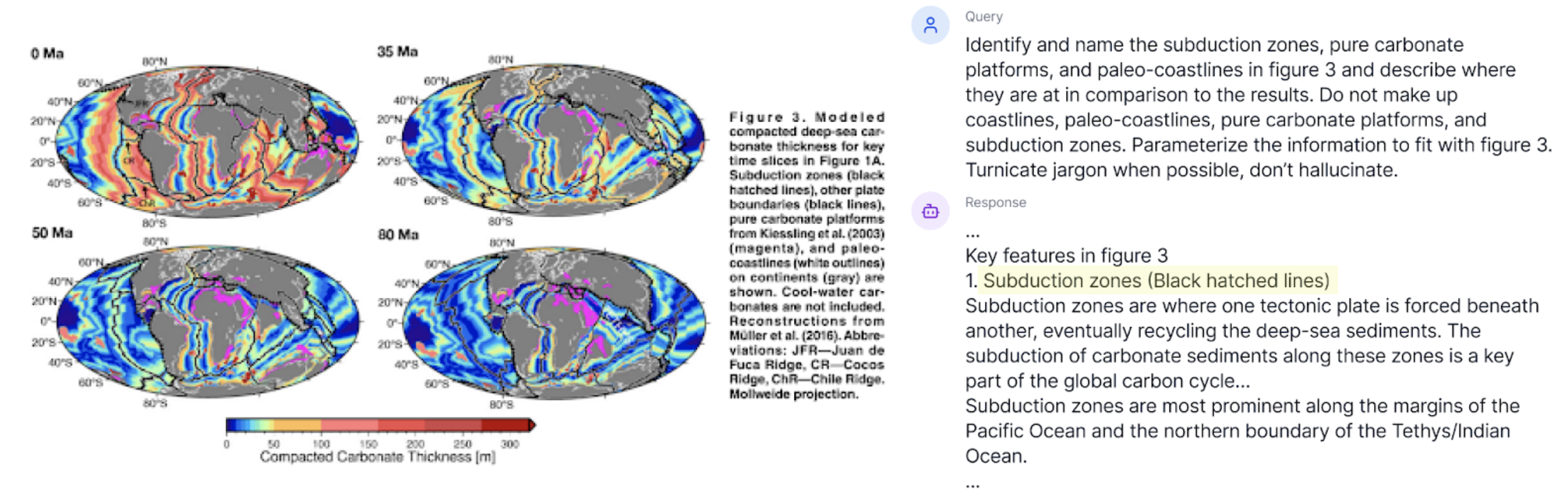}
\Description{On the left is Figure 3 from the Sequestration and Subduction paper, along with the original caption. The figure shows four models of the world with different colors indicating compacted carbonate thickness and black lines indicating subduction zones. On the right is the original user query asking the AI to "Identify and name the subduction zones, pure carbonate platforms, and paelo-coastlines...", and below that is the model response where it starts off saying the subduction zones are "Black hatched lines" (highlighted in yellow to indicate the point of user confusion) and then proceeds to explain what a subduction zone is.}
\caption{AI descriptions provided visual information, but did not help with interpretation: BLV3 found no use from the AI response telling her subduction zones were represented by black lines. Instead she wanted to know where they were spatially, and along which specific countries. She was querying against the figure shown here from the paper ``Sequestration and subduction of deep-sea carbonate in the global ocean since the Early Cretaceous'' by \citet{Subduction_Dutkiewicz}.}
\label{SubductionZones}
\end{figure*}

\begin{figure*}[h]
\centering
\includegraphics[width=\textwidth]{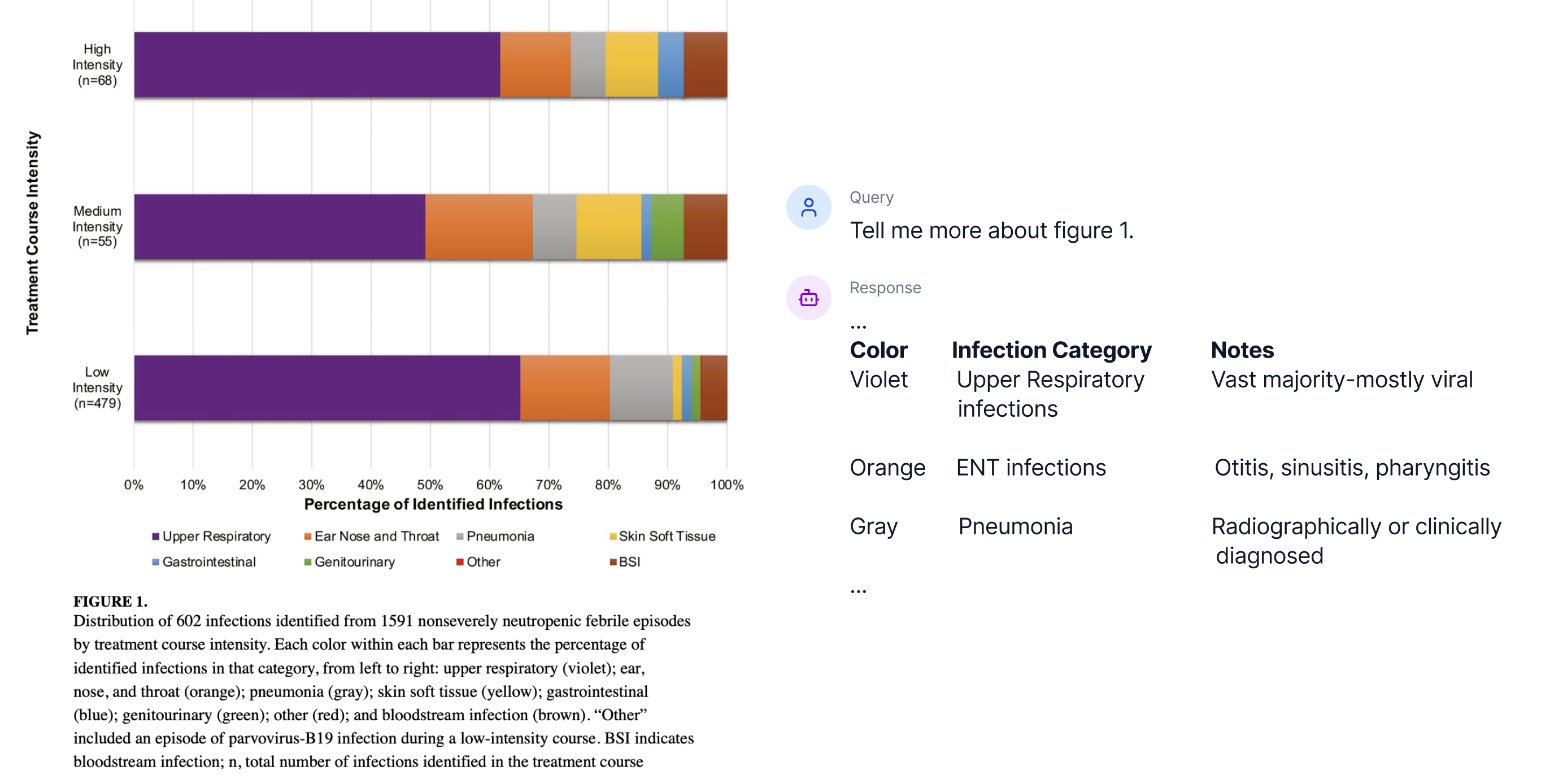}
\Description{On the left is Figure 1 from the Infectious events in pediatric patients paper, with the original caption. Figure 1 shows a bar chart with different colors (purple, orange, gray, yellow, etc.) that represent different types of infections. On the right is the user query to the AI, saying "Tell me more about figure 1", and below that is the AI response. The portion of the response that is shown presents a partial table, where the AI has presented the bar color (e.g., Violet) along with the infection category it represents (e.g., Upper Respiratory) and additional notes (e.g., Vast majority-mostly viral).}
\caption{AI descriptions providing visual information but no interpretation also frustrated sighted participants. S2 expected the AI system to do more interpretation by providing percentages and comparisons of the different infections, and less direct translation of visual semantics (\textit{i.e.}, the colors of the bars) for this chart from ``Infectious events in pediatric patients with acute lymphoblastic leukemia/lymphoma undergoing evaluation for fever without severe neutropenia'' by \citet{patel2022infectious}.}
\label{S2_ColorBarChart}
\end{figure*}

Another recurring theme was the friction between being told visual semantics of a figure, such as line colors, versus its scientific significance. BLV3 noted that being told \textit{``subduction zones are shown by black lines''} in an AI response was useless without being told where those zones were located spatially (Figure \ref{SubductionZones}). Sighted participants shared some of these frustrations, with S2 expecting the AI to \textit{``know what percentage are each of the color''} in a bar chart rather than merely reciting the caption and colors of the bars (Figure \ref{S2_ColorBarChart}). BLV researchers had additional motivation to understand scientific significance of figures to establish \textit{``common ground''} with sighted collaborators (BLV1) and to recover details not represented in the main text. Depending on the domain, this could be even more critical---S1 noted that in his field, materials science, papers are traditionally written under the assumption that the reader is \textit{``staring at the figures''} as they read. BLV5, a neuroscience and computer science researcher, described his optimal AI description flow involving both interpretation and a pure visual description of figures:

\begin{quote}
    \textit{``I would start by getting the interpretation, and then you get to `okay, but how does the image actually look like?' Because first I have to understand what the image is because then I generate a visual model of that image in my brain; and then asking how does this image actually look like I can adapt this visual model that I have to to what in reality the image looks like.''}---\textbf{BLV5}
\end{quote}

For BLV participants, another key point of struggle was when AI responses left details out in a manner that felt arbitrary. We observed one example of such a failure case where AI descriptions of multi-part figures summarized only a subset of sub-panels. As a result, BLV1 interpreted the figure as having fewer components, since additional panels were omitted from both the AI output and the paper text. This omission led to confusion, as those sub-parts were inaccessible to BLV1. BLV participants also reported that dense, text-based descriptions of visual content were difficult to parse and reduced the helpfulness of AI responses. BLV1 noted that is responses become too cumbersome, he would \textit{``just ask a colleague''} instead. Participants also suggested alternative formats that could better support interpretation, including interactive visualizations (BLV1) or \textit{``timeline-style summaries''} (BLV4) that can help break down the sequential structure of figures. BLV2 also described wanting to using AI to generate tactile graphics for blind students, highlighting interest in representations beyond text descriptions.

\subsection{Validation, Trust, and Privacy}\label{ValidationTrustPrivacy}
Finally, we present how participants validate responses from QA systems, and where loss of trust and privacy become concerns that hinder the continued use of such tools.

\textbf{Validation across QA agent and source material.}
Since participants first used each QA tool with both familiar and unfamiliar papers without access to the source document, we could observe siloed interactions with the system before they referenced the paper later. For several participants, including BLV4 and S1, the inability to validate AI outputs against the paper initially led them to say they would stop using the tool, either on its own or entirely. In the study, this point marked the transition to the next tool condition, where participants used the subsequent QA system with access to the original PDF on their device. For many, the AI served as a directional aid: BLV1 and S4 used it to identify relevant figures or mechanisms before returning to the paper to validate the response. This cross-checking often followed perceived inaccuracies---for example, S1 manually inspected figures from the original paper after receiving what he suspected was incorrect information about a test setup. S3 emphasized the need for the AI tools to be more reliable, stating that she would not typically open the paper during QA interaction unless explicitly verifying a suspected hallucination. Notably, when participants did reference the paper, none of the BLV participants attempted to use figure alt-text as part of their validation or cross-checking; they instead relied on the caption and the portions of text from the same section as the figure.

\textbf{Loss of trust.}
We found there was extreme fragility with participants' trust in AI systems for science QA---as S1 described, \textit{``I can't fact check everything, so if I know it's missed something, it's like a one strike thing.''} For researchers in high-precision fields like materials science, even \textit{``very small changes... imply so much different chemically''} (S1). This intolerance for error was echoed by BLV3, who noted that while an AI might be instructed to make a guess when uncertain, \textit{``in science you don't do that.''} She expressed how the stakes of these hallucinations are even higher for BLV participants: 

\begin{quote}
    \textit{``For a sighted person it would be easy for them to see if AI hallucinated and made up facts or numbers; but as a blind person I would actually need an accessible version of the document to actually determine whether or not it's been hallucinated.''}---\textbf{BLV3}
\end{quote}

We also observed that when the AI provided an inaccurate response, it affected the researchers' confidence, with S4 noting that a flawed response made her \textit{``slightly question [her] own understanding of the paper,''} at which point she would move away from the agent entirely to consult the manuscript. Participants additionally expressed frustration when they could not distinguish between the authors' original findings and AI interpretations. BLV2 suggested this might be due to a lack of tone, noting that unlike a human colleague, the AI made it difficult to discern \textit{``when they [AI] are synthesizing their own ideas versus when it's not new ideas.''}

\textbf{Privacy considerations.}
While the majority of participants did not express concerns about the privacy implications of their queries, one BLV participant, BLV2, raised this as a significant issue, preferring not to tell AI systems that she cannot see a figure: 

\begin{quote}
    \textit{``The number of AI systems who say that `We're not using your information for X' and then actually have been using your information for X all along...  I usually try creatively wording it: `Oh, it's unclear what's going on here.' ''}---\textbf{BLV2}
\end{quote}

She adopted this strategy during the study after receiving an unhelpful response to her initial query, \textit{``Please summarize the 1st flow chart in this document.''} Contrastingly, BLV5 as another BLV scientist very explicitly queried for \textit{``blind-friendly''} content as a part of his QA interactions, and had no concern about AI systems inferring his blindness: 

\begin{quote}
    \textit{``Generate a blind friendly, math and code focused of the figures of the paper. Explain to me how they were generated, how would I present them on a lab meeting and how I could generate similar visuals.''}---\textbf{BLV5}, querying an unfamiliar paper using Gemini
\end{quote}

These contrasting perspectives underscore that privacy concerns among BLV scientists are not uniform, but instead shaped by individual preferences and interaction goals.

\section{Discussion}
We now reflect on findings from our study and present implications for future inclusive and flexible QA systems. We close by discussing limitations of the present work and opportunities for future work.

\subsection{Reflecting on Multimodal QA in Scientific Workflows}
In this work, we asked the research question: how do scientists with diverse visual access needs query multimodal scientific papers using AI systems, and what practices and preferences shape this interaction? Our findings provide a holistic view of how BLV and sighted scientists use multimodal QA systems with papers, revealing shared sensemaking practices and differences shaped by visual access. Both groups started with overview questions to situate themselves in the paper before pivoting to ask about specific figures or methods. This suggests that while access needs differ, the underlying workflow of scientific sensemaking is consistent across BLV and sighted scientists. Related to sensemaking, another common expectation across both BLV and sighted scientists was that QA systems should support figure and table \textit{interpretation}, not merely provide surface-level descriptions of visual semantics. Often when systems did provide visual descriptions, they emphasized the wrong details, overlooking scientifically relevant features such as graph axes, quantitative values, or spatial relationships.

These shared expectations coexisted with distinct patterns in how participants used and benefited from multimodal QA systems. For BLV participants, QA systems functioned as a primary access layer for visual content, with almost half of their queries (49\%) directed toward specific figures and tables to recover chart axes, data relationships, and structural details. In contrast, sighted participants treated QA systems as an efficiency layer, using them more to synthesize methods and findings (56\%). This divergence suggests that multimodal QA systems serve different roles in scientist workflows---for BLV scientists, they act as an interpretive access layer for visual content that is otherwise challenging to access, whereas for sighted scientists, they streamline synthesis and navigation of already-accessible information.

We further observed differences in how participants establish and verify knowledge. Sighted participants can more easily validate QA systems by cross-checking against the source material to identify mismatches between figures and descriptions, as S1 did when he reviewed information he believed was inaccurate. In contrast, BLV participants often lacked the means to independently validate AI outputs, because, as BLV3 highlighted, the source material itself is often inaccessible. This creates a gap in agency, where BLV scientists must rely more heavily on AI-generated interpretations without the ability to properly verify them, consult with a sighted friend or colleague, or come up with workarounds that are time-consuming such as producing tactile-printed graphics or code representations of the scientific figures. Prior accessibility research has emphasized the importance of independent access and verification in accessibility technologies \cite{Wobbrock_AbilityBasedDesign, Alharbi_MisfittingWithAI, Tang_EverydayUncertainty}; our findings show that these concerns remain just as critical in AI-mediated scientific workflows, where the use of such QA tools can amplify existing asymmetries between BLV and sighted scientists.

At the same time, participants’ practices revealed opportunities for inclusive design that benefit a broader range of users. For instance, BLV1 and BLV5 translated figures into code or mathematical representations to obtain more accessible formats, while a sighted participant, S1, similarly reported manually tabulating data from graphs when reviewing papers. These parallels suggest that representations motivated by accessibility needs may also support broader scientific QA workflows \cite{Blackwell2016}, echoing prior findings in productivity support \cite{Abdolrahmani_BlindLeadingSighted} and accessible visualization \cite{Li_GeoVisA11y}.

\subsection{Implications for Flexible, Inclusive QA Systems}
We present guidelines for QA systems to provide more inclusive and effective multimodal experiences for BLV and sighted scientists.

\textbf{1. Provide domain-aware, interpretive figure descriptions.} Even when current AI systems correctly describe \textit{what} appears in a figure (\textit{e.g.}, colors, lines, continents), they can fail to explain \textit{why} those elements matter or to extract the scientifically relevant details. This is problematic for both BLV and sighted scientists---BLV3 could not interpret the locations of subduction zones from a figure description, while S2 was told only the colors of chart bars rather than the underlying percentages of diseases. QA agents must provide domain-aware, interpretive figure descriptions that prioritize scientific meaning over literal visual description alone.

\textbf{2. Support ``beyond text'' multimodal representations.} Our findings suggest that text alone may be insufficient for representing complex scientific visuals. BLV and sighted participants described wanting figures translated into other forms, including mathematical expressions, code, tables, and more structured breakdowns that better supported analysis. Some BLV scientists also wanted progressive or sequential explanations of images, rather than a single static description. QA responses should support richer, more flexible representations of figures that extend beyond text.

\textbf{3. Design for progressive scientific scaffolding.} Both BLV and sighted scientists followed a similar ``funnel'' strategy, beginning with broad orienting queries before narrowing to specific figures, methods, or findings. Current QA interfaces do not support this progression, requiring users to manage that structure themselves. QA systems should instead scaffold responses from overview to detail while preserving the broader scientific context throughout. For BLV scientists in particular, this should include upfront figure and table overviews, as they often constructed these themselves in our studies for quick multimodal content understanding.

\textbf{4. Provide increased transparency around synthesis versus quoting source material.} Trust in scientific QA is fragile, particularly when users cannot tell whether a response is directly grounded in the paper or synthesized by the model. BLV2 had difficulty distinguishing when the AI was \textit{``synthesizing their own ideas''} versus reflecting the paper itself. Scientific QA systems should clearly signal whether information is quoted, summarized, or inferred, so users can better assess the validity of a response. This is especially important for BLV scientists who expressed having a harder time verifying potential hallucinations and false claims.

\textbf{5. Support accessibility without requiring disclosure.} Our findings revealed a tension between wanting more blind-accessible outputs and not wanting to explicitly disclose blindness to the system. While BLV5 directly requested blind-oriented responses, BLV2 used indirect phrasing to elicit better descriptions without revealing disability status. QA systems should therefore support accessibility features in flexible, on-demand ways that do not depend on persistent disclosure or user profiling. More broadly, systems should avoid treating these supports as only relevant to blind users, since our study shows that clearer detail, more explicit interpretation, and diverse representations such as code and math versions of figures may benefit a wider range of scientists.

\subsection{Limitations and Future Work}

This work has three main limitations. First, our participant pool was small (five BLV and five sighted), and though we recruited participants from several different domains, this smaller pool limits the generalizability of our findings. Second, all of the BLV participants were PhD students; their experiences may differ from more senior scientists, who may have developed different strategies for engaging with figures, tables, and AI tools over time. Finally, participants worked with domain-specific papers of their choosing. While this captured naturalistic engagement with content from their field, it limited direct comparison across participants in how they interpret and respond to a consistent set of complex figures.

In addition to addressing the limitations listed above, future work should consider expanding beyond STEM domains, where figures and tables may take different forms and play different roles in sensemaking. Exploring disciplines such as the social sciences or humanities may reveal alternative practices and needs for multimodal QA systems. Future systems could also move toward deeper personalization, incorporating knowledge of a scientist’s research area, prior work, and preferences to provide more contextually relevant responses and recommendations. Finally, while exploring tactile and spatial understanding of scientific figures was out of scope for the present work as we focused on QA systems, future work could examine how spatial reasoning shapes scientific figure interpretation and whether haptic or tactile displays can better support validation of multimodal QA responses, particularly for figures where text descriptions alone may be insufficient.

Our dataset contribution of 115 queries and responses by BLV and sighted scientists may also support future work beyond this study. Prior work has shown the value of such resources: \citet{Kim_ChartQA} dataset of 979 BLV users’ questions about visualizations has informed later systems work, including the \textit{VizAbility} conversational chart-access system \cite{Gorniak_VizAbility}. Similarly, we hope our dataset of scientist-authored queries and responses for multimodal scientific documents can help ground future benchmarking, interaction design, and evaluation in real user practices and accessibility needs.

\section{Conclusion}
In this work, we conducted interviews with five BLV and five sighted scientists across different STEM domains to understand how scientists with diverse visual access needs query multimodal scientific papers using AI systems, and what practices and preferences shape this interaction. We further contribute the set of queries and responses from our participants' QA interactions with two systems, ChatGPT and Gemini. As AI-powered systems are increasingly used for productivity support, it is critical that scientific access not be treated as a secondary concern, but as a core requirement of how these tools are designed. By understanding both the shared and divergent ways BLV and sighted scientists use these systems, such as bridging access to visual content and accelerating synthesis and verification, and where current AI falls short of scientific precision, we can better shape QA systems that support rigorous and inclusive scientific work.

\begin{acks}
We thank Matt Latzke and Evie Cheng for their help with this work.
\end{acks}

\bibliographystyle{ACM-Reference-Format}
\bibliography{references}

\clearpage
\onecolumn
%TC:ignore
\appendix
\section{115 Queries by BLV and Sighted Participants}\label{115_Queries}
We present the 115 queries from the participants below, along with the query type, paper title, link, which tool they used, and whether the paper was familiar or unfamiliar to them. Each query type is numbered based on the category we identified for that query: (1) General or Overview, (2) Figures or Tables, (3) Methods or Findings, and (4) Miscellaneous. The full set of responses along with these queries can be found in the supplementary materials.

\begin{longtable}{p{0.8cm} p{3.4cm} p{0.8cm} p{1.2cm} p{1cm} p{3.4cm} p{3.4cm}}
\caption{Participant Queries and Paper Metadata} \label{tab:queries} \\
\toprule
\textbf{ID} & \textbf{Query} & \textbf{Type} & \textbf{Familiar?} & \textbf{Tool} & \textbf{Paper Title} & \textbf{Paper Link} \\ \midrule
\endfirsthead

\multicolumn{7}{c}%
{{\bfseries \tablename\ \thetable{} -- continued from previous page}} \\
\midrule
\textbf{ID} & \textbf{Query} & \textbf{Type} & \textbf{Familiar?} & \textbf{Tool} & \textbf{Paper Title} & \textbf{Paper Link} \\ \midrule
\endhead

% \midrule
% \multicolumn{7}{r}{{Continued on next page...}} \\
% \bottomrule
% \endfoot

\bottomrule
\endlastfoot
BLV1 & List and summarize all the figures in this research paper. Include the provided alt text description (if any) and the title/caption in the figure. & 2 & Familiar & ChatGPT & Typing Haptically: Towards Enabling Non-auditory Smartphone Text Entry with Haptic Feedback for Blind and Low Vision Users & \url{http://dl.acm.org/doi/10.1145/3746059.3747801} \\
BLV1 & In figure 9, describe the type of chart and the chart itself. & 2 & Familiar & ChatGPT & Typing Haptically: Towards Enabling Non-auditory Smartphone Text Entry with Haptic Feedback for Blind and Low Vision Users & \url{http://dl.acm.org/doi/10.1145/3746059.3747801} \\
BLV1 & go through each individual participant plotted and describe the individual trends. Try to estimate the percentages if not explicitly marked on the figure. & 2 & Familiar & ChatGPT & Typing Haptically: Towards Enabling Non-auditory Smartphone Text Entry with Haptic Feedback for Blind and Low Vision Users & \url{http://dl.acm.org/doi/10.1145/3746059.3747801} \\
BLV1 & Give me an overview of Figure 3. Include the original caption and alt text (if any) and provide an overall description of the figure. & 2 & Familiar & Gemini & Typing Haptically: Towards Enabling Non-auditory Smartphone Text Entry with Haptic Feedback for Blind and Low Vision Users & \url{http://dl.acm.org/doi/10.1145/3746059.3747801} \\
BLV1 & for each category of sounds, describe the corresponding haptic design based on the details in the figure and other text from the paper. & 3 & Familiar & Gemini & Typing Haptically: Towards Enabling Non-auditory Smartphone Text Entry with Haptic Feedback for Blind and Low Vision Users & \url{http://dl.acm.org/doi/10.1145/3746059.3747801} \\
BLV1 & Give me an overview of this paper. Tell me what the contribution(s) are and an overall summary of the structure of the paper. & 1 & Unfamiliar & Gemini & MagnePins: A Modular, Affordable, and DIY Refreshable Braille and Tactile Display & \url{https://dl.acm.org/doi/10.1145/3746059.3747692} \\
BLV1 & what's the stated cost of building this display? & 3 & Unfamiliar & Gemini & MagnePins: A Modular, Affordable, and DIY Refreshable Braille and Tactile Display & \url{https://dl.acm.org/doi/10.1145/3746059.3747692} \\
BLV1 & In list item 3 that you mentioned, dive deeper into the technical details of how the pins are actuated and the modular design. & 3 & Unfamiliar & Gemini & MagnePins: A Modular, Affordable, and DIY Refreshable Braille and Tactile Display & \url{https://dl.acm.org/doi/10.1145/3746059.3747692} \\
BLV1 & List the figures (if any) that illustrate the actuation mechanism/design. & 2 & Unfamiliar & Gemini & MagnePins: A Modular, Affordable, and DIY Refreshable Braille and Tactile Display & \url{https://dl.acm.org/doi/10.1145/3746059.3747692} \\
BLV1 & Tell me about all the figures that illustrate how the actuation mechanism works in the tool described in this paper. Don't list figures that detail the fabrication process of how the researchers built the device, but do list the figures that describe how the actuation works. & 2 & Unfamiliar & ChatGPT & MagnePins: A Modular, Affordable, and DIY Refreshable Braille and Tactile Display & \url{https://dl.acm.org/doi/10.1145/3746059.3747692} \\
BLV1 & Describe how the actuation mechanism works using figure 6 and details included in section 4.3. How do the pins stay raised and how do they get lowered if there isn't an actuator for every pin as mentioned in other parts of the paper & 2 & Unfamiliar & ChatGPT & MagnePins: A Modular, Affordable, and DIY Refreshable Braille and Tactile Display & \url{https://dl.acm.org/doi/10.1145/3746059.3747692} \\
S1 & describe the main sealing methods tested in this paper and describe the main test conditions for the shaft seal & 3 & Familiar & ChatGPT & High Temperature Braided Rope Seals for Static Sealing Applications & \url{https://arc.aiaa.org/doi/10.2514/2.5219} \\
S1 & show a diagram & 4 & Familiar & ChatGPT & High Temperature Braided Rope Seals for Static Sealing Applications & \url{https://arc.aiaa.org/doi/10.2514/2.5219} \\
S1 & tell me about the shaft seal performance at high temperature and high compression & 3 & Familiar & ChatGPT & High Temperature Braided Rope Seals for Static Sealing Applications & \url{https://arc.aiaa.org/doi/10.2514/2.5219} \\
S1 & what was the shape of the relationship of leak rate vs pressure & 3 & Familiar & ChatGPT & High Temperature Braided Rope Seals for Static Sealing Applications & \url{https://arc.aiaa.org/doi/10.2514/2.5219} \\
S1 & tabulate figure 3 & 2 & Familiar & Gemini & High Temperature Braided Rope Seals for Static Sealing Applications & \url{https://arc.aiaa.org/doi/10.2514/2.5219} \\
S1 & my shaft diameter is 3" with differential pressure of 8 psi. What would my leak rate be in Standard L/min & 3 & Familiar & Gemini & High Temperature Braided Rope Seals for Static Sealing Applications & \url{https://arc.aiaa.org/doi/10.2514/2.5219} \\
S1 & any authors still alive and practicing & 4 & Familiar & Gemini & High Temperature Braided Rope Seals for Static Sealing Applications & \url{https://arc.aiaa.org/doi/10.2514/2.5219} \\
S1 & summarize the paper & 1 & Unfamiliar & ChatGPT & Effect of Heat Treatment on Thermal Expansion Behavior and Corrosion Resistance of Martensitic Stainless Steel Manufactured by Submerged Arc Welding & \url{https://www.semanticscholar.org/paper/Effect-of-Heat-Treatment-on-Thermal-Expansion-and-Wang/3a0fa050f580d71f374f175bfddf57942a229029} \\
S1 & How does tempering effect CTE? & 3 & Unfamiliar & ChatGPT & Effect of Heat Treatment on Thermal Expansion Behavior and Corrosion Resistance of Martensitic Stainless Steel Manufactured by Submerged Arc Welding & \url{https://www.semanticscholar.org/paper/Effect-of-Heat-Treatment-on-Thermal-Expansion-and-Wang/3a0fa050f580d71f374f175bfddf57942a229029} \\
S1 & How much residual austenite is in annealed 410 compared to H13 & 3 & Unfamiliar & ChatGPT & Effect of Heat Treatment on Thermal Expansion Behavior and Corrosion Resistance of Martensitic Stainless Steel Manufactured by Submerged Arc Welding & \url{https://www.semanticscholar.org/paper/Effect-of-Heat-Treatment-on-Thermal-Expansion-and-Wang/3a0fa050f580d71f374f175bfddf57942a229029} \\
S1 & would this paper be helpful for assessing the effect of annealing on 410 stainless? & 3 & Unfamiliar & ChatGPT & Effect of Heat Treatment on Thermal Expansion Behavior and Corrosion Resistance of Martensitic Stainless Steel Manufactured by Submerged Arc Welding & \url{https://www.semanticscholar.org/paper/Effect-of-Heat-Treatment-on-Thermal-Expansion-and-Wang/3a0fa050f580d71f374f175bfddf57942a229029} \\
S1 & tell me how this paper relates to the annealed cte of 410 stainless compared to full hard & 3 & Unfamiliar & Gemini & Effect of Heat Treatment on Thermal Expansion Behavior and Corrosion Resistance of Martensitic Stainless Steel Manufactured by Submerged Arc Welding & \url{https://www.semanticscholar.org/paper/Effect-of-Heat-Treatment-on-Thermal-Expansion-and-Wang/3a0fa050f580d71f374f175bfddf57942a229029} \\
S1 & what are the reciprocal lattice parameters of two carbides & 3 & Unfamiliar & Gemini & Effect of Heat Treatment on Thermal Expansion Behavior and Corrosion Resistance of Martensitic Stainless Steel Manufactured by Submerged Arc Welding & \url{https://www.semanticscholar.org/paper/Effect-of-Heat-Treatment-on-Thermal-Expansion-and-Wang/3a0fa050f580d71f374f175bfddf57942a229029} \\
S1 & in figure 6 list the major 2 theta peaks & 2 & Unfamiliar & Gemini & Effect of Heat Treatment on Thermal Expansion Behavior and Corrosion Resistance of Martensitic Stainless Steel Manufactured by Submerged Arc Welding & \url{https://www.semanticscholar.org/paper/Effect-of-Heat-Treatment-on-Thermal-Expansion-and-Wang/3a0fa050f580d71f374f175bfddf57942a229029} \\
S1 & what are they doing with the electro-chem setup & 3 & Unfamiliar & Gemini & Effect of Heat Treatment on Thermal Expansion Behavior and Corrosion Resistance of Martensitic Stainless Steel Manufactured by Submerged Arc Welding & \url{https://www.semanticscholar.org/paper/Effect-of-Heat-Treatment-on-Thermal-Expansion-and-Wang/3a0fa050f580d71f374f175bfddf57942a229029} \\
S1 & what was the result and how did phase fraction drive them & 3 & Unfamiliar & Gemini & Effect of Heat Treatment on Thermal Expansion Behavior and Corrosion Resistance of Martensitic Stainless Steel Manufactured by Submerged Arc Welding & \url{https://www.semanticscholar.org/paper/Effect-of-Heat-Treatment-on-Thermal-Expansion-and-Wang/3a0fa050f580d71f374f175bfddf57942a229029} \\
S2 & Give me a summary of this paper & 1 & Familiar & Gemini & Infectious events in pediatric patients with acute lymphoblastic leukemia/lymphoma undergoing evaluation for fever without severe neutropenia & \url{https://pubmed.ncbi.nlm.nih.gov/36238979/} \\
S2 & what are some future directions for this work & 4 & Familiar & Gemini & Infectious events in pediatric patients with acute lymphoblastic leukemia/lymphoma undergoing evaluation for fever without severe neutropenia & \url{https://pubmed.ncbi.nlm.nih.gov/36238979/} \\
S2 & Give me a summary of this paper & 1 & Familiar & ChatGPT & Infectious events in pediatric patients with acute lymphoblastic leukemia/lymphoma undergoing evaluation for fever without severe neutropenia & \url{https://pubmed.ncbi.nlm.nih.gov/36238979/} \\
S2 & tell me about figure 1 & 2 & Familiar & ChatGPT & Infectious events in pediatric patients with acute lymphoblastic leukemia/lymphoma undergoing evaluation for fever without severe neutropenia & \url{https://pubmed.ncbi.nlm.nih.gov/36238979/} \\
S2 & give me a breakdown of each type of infection and frequencies for each course & 3 & Familiar & ChatGPT & Infectious events in pediatric patients with acute lymphoblastic leukemia/lymphoma undergoing evaluation for fever without severe neutropenia & \url{https://pubmed.ncbi.nlm.nih.gov/36238979/} \\
S2 & Give me a summary of this paper as I am not familiar with it & 1 & Unfamiliar & ChatGPT & Invasive fungal diseases impact on outcome of childhood ALL – an analysis of the international trial AIEOP-BFM ALL 2009 & \url{https://pubmed.ncbi.nlm.nih.gov/36509893/} \\
S2 & How many patients had proven vs probable infection? & 3 & Unfamiliar & ChatGPT & Invasive fungal diseases impact on outcome of childhood ALL – an analysis of the international trial AIEOP-BFM ALL 2009 & \url{https://pubmed.ncbi.nlm.nih.gov/36509893/} \\
S2 & whats the breakdown of organism within the two groups? & 3 & Unfamiliar & ChatGPT & Invasive fungal diseases impact on outcome of childhood ALL – an analysis of the international trial AIEOP-BFM ALL 2009 & \url{https://pubmed.ncbi.nlm.nih.gov/36509893/} \\
S2 & What are the breakdowns of type of fungal infection within the proven and probable categories of infection? & 3 & Unfamiliar & Gemini & Invasive fungal diseases impact on outcome of childhood ALL – an analysis of the international trial AIEOP-BFM ALL 2009 & \url{https://pubmed.ncbi.nlm.nih.gov/36509893/} \\
S2 & Can you give me a summary of figure 2? & 2 & Unfamiliar & Gemini & Invasive fungal diseases impact on outcome of childhood ALL – an analysis of the international trial AIEOP-BFM ALL 2009 & \url{https://pubmed.ncbi.nlm.nih.gov/36509893/} \\
S3 & Can you summarize the statistically significant outcomes from the fasting intervention & 3 & Familiar & Gemini & Fasting mimicking diet during neo-adjuvant chemotherapy in breast cancer patients: a randomized controlled trial study & \url{https://pubmed.ncbi.nlm.nih.gov/39703333/} \\
S3 & What was the size of the study and any other details of the population? & 3 & Familiar & Gemini & Fasting mimicking diet during neo-adjuvant chemotherapy in breast cancer patients: a randomized controlled trial study & \url{https://pubmed.ncbi.nlm.nih.gov/39703333/} \\
S3 & What are the specifics of the intervention? & 3 & Familiar & Gemini & Fasting mimicking diet during neo-adjuvant chemotherapy in breast cancer patients: a randomized controlled trial study & \url{https://pubmed.ncbi.nlm.nih.gov/39703333/} \\
S3 & what are the significant outcomes from this intervention & 3 & Familiar & ChatGPT & Fasting mimicking diet during neo-adjuvant chemotherapy in breast cancer patients: a randomized controlled trial study & \url{https://pubmed.ncbi.nlm.nih.gov/39703333/} \\
S3 & what is the most meangingful metric from table 2? & 2 & Familiar & ChatGPT & Fasting mimicking diet during neo-adjuvant chemotherapy in breast cancer patients: a randomized controlled trial study & \url{https://pubmed.ncbi.nlm.nih.gov/39703333/} \\
S3 & i'm interested in clinical data for plant polyphenols, does this study reference any clinical studies & 3 & Unfamiliar & Gemini & Plant Polyphenols as Heart's Best Friends: From Health Properties, to Cellular Effects, to Molecular Mechanisms of Action & \url{https://pubmed.ncbi.nlm.nih.gov/39940685/} \\
S3 & can you share more about the specific clinical outcomes for curcumin & 3 & Unfamiliar & Gemini & Plant Polyphenols as Heart's Best Friends: From Health Properties, to Cellular Effects, to Molecular Mechanisms of Action & \url{https://pubmed.ncbi.nlm.nih.gov/39940685/} \\
S3 & Are there specific mechanisms of action described that apply to all plant polyphenols discussed? & 3 & Unfamiliar & Gemini & Plant Polyphenols as Heart's Best Friends: From Health Properties, to Cellular Effects, to Molecular Mechanisms of Action & \url{https://pubmed.ncbi.nlm.nih.gov/39940685/} \\
S3 & Are there any cancer studies mentioned? & 3 & Unfamiliar & Gemini & Plant Polyphenols as Heart's Best Friends: From Health Properties, to Cellular Effects, to Molecular Mechanisms of Action & \url{https://pubmed.ncbi.nlm.nih.gov/39940685/} \\
S3 & does this paper differentiate between dietary sources of polyphenols and supplements and their difference in effectiveness? & 3 & Unfamiliar & Gemini & Plant Polyphenols as Heart's Best Friends: From Health Properties, to Cellular Effects, to Molecular Mechanisms of Action & \url{https://pubmed.ncbi.nlm.nih.gov/39940685/} \\
S3 & can you dig into this further? Is there any data demonstrating dietary polyphenols with clinical outcomes & 3 & Unfamiliar & Gemini & Plant Polyphenols as Heart's Best Friends: From Health Properties, to Cellular Effects, to Molecular Mechanisms of Action & \url{https://pubmed.ncbi.nlm.nih.gov/39940685/} \\
S3 & Are you sure this was a study of diet and not supplement/extract ? Blood pressure — RCTs and meta-analyses: Several RCTs and meta-analyses show dietary polyphenols (epicatechin, catechins, anthocyanins, quercetin) lower systolic and diastolic blood pressure (effects often dose-dependent; e.g., epicatechin minimum \textbackslash{}textasciitilde{}25 mg/day). & 3 & Unfamiliar & Gemini & Plant Polyphenols as Heart's Best Friends: From Health Properties, to Cellular Effects, to Molecular Mechanisms of Action & \url{https://pubmed.ncbi.nlm.nih.gov/39940685/} \\
S3 & in one table animal models given doxorubicin were summarized, can you tell me the outcomes in a simple summary? & 2 & Unfamiliar & Gemini & Plant Polyphenols as Heart's Best Friends: From Health Properties, to Cellular Effects, to Molecular Mechanisms of Action & \url{https://pubmed.ncbi.nlm.nih.gov/39940685/} \\
S3 & in table 7 what is the overall takeaway of resveratrol in animal models? & 2 & Unfamiliar & Gemini & Plant Polyphenols as Heart's Best Friends: From Health Properties, to Cellular Effects, to Molecular Mechanisms of Action & \url{https://pubmed.ncbi.nlm.nih.gov/39940685/} \\
BLV2 & What conclusions do the authors come up with regarding screen vs text reading with a list of caveats included? & 3 & Unfamiliar & ChatGPT & A Comparison of Children’s Reading on Paper Versus Screen: A Meta-Analysis & \url{https://www.semanticscholar.org/paper/A-Comparison-of-Children%E2%80%99s-Reading-on-Paper-Versus-Furenes-Kucirkova/d3783034db2c17095e713a0a801d01a031fbe58f} \\
BLV2 & Please summarize the 1st flow chart in this document & 2 & Unfamiliar & Gemini & A Comparison of Children’s Reading on Paper Versus Screen: A Meta-Analysis & \url{https://www.semanticscholar.org/paper/A-Comparison-of-Children%E2%80%99s-Reading-on-Paper-Versus-Furenes-Kucirkova/d3783034db2c17095e713a0a801d01a031fbe58f} \\
BLV2 & It's unclear to me what is going on in figure one. Could you please provide a high level overview of the entire chart setup & 2 & Unfamiliar & Gemini & A Comparison of Children’s Reading on Paper Versus Screen: A Meta-Analysis & \url{https://www.semanticscholar.org/paper/A-Comparison-of-Children%E2%80%99s-Reading-on-Paper-Versus-Furenes-Kucirkova/d3783034db2c17095e713a0a801d01a031fbe58f} \\
BLV2 & Give me an overview of the background of author Rule from internet data including background on research with blind students such as in this article & 4 & Familiar & ChatGPT & Tactile Earth and Space Science Materials for Students with Visual Impairments: Contours, Craters, Asteroids, and Features of Mars & \url{https://www.tandfonline.com/doi/abs/10.5408/1.3651404} \\
BLV2 & Is there an image in the paper that would appear to show one of the Mars maps and if so what sort of information seems to be highlighted within the image & 2 & Familiar & ChatGPT & Tactile Earth and Space Science Materials for Students with Visual Impairments: Contours, Craters, Asteroids, and Features of Mars & \url{https://www.tandfonline.com/doi/abs/10.5408/1.3651404} \\
BLV2 & Using your knowledge reagrding Vallis Marineris and ideas of tactile graphics what would you suggest highlighting for understanding for blind students? Compare back to what the authors did with analysis of match percentage to your answer & 3 & Familiar & ChatGPT & Tactile Earth and Space Science Materials for Students with Visual Impairments: Contours, Craters, Asteroids, and Features of Mars & \url{https://www.tandfonline.com/doi/abs/10.5408/1.3651404} \\
BLV3 & Summarize the findings in this uploaded paper by Dutkiewicz et al, focus on the results, figures, diagrams, graphs, and images. If there are no visuals such as figures, diagrams, graphs, or images then do not hallucinate them. Keep the summary concise, include any math mentioned in the paper. & 1 & Unfamiliar & ChatGPT & Sequestration and subduction of deep-sea carbonate in the global ocean since the Early Cretaceous & \url{https://pubs.geoscienceworld.org/gsa/geology/article-abstract/47/1/91/567642/Sequestration-and-subduction-of-deep-sea-carbonate} \\
BLV3 & What was in figures 1a and 1b, and why did you mention the fourth bulletin point under 1. Carbonate Sequestration and Storage as you have in parenthesis that this is covering figures 1C and 1D yet mention 1B explain in concise details do not make up information & 2 & Unfamiliar & ChatGPT & Sequestration and subduction of deep-sea carbonate in the global ocean since the Early Cretaceous & \url{https://pubs.geoscienceworld.org/gsa/geology/article-abstract/47/1/91/567642/Sequestration-and-subduction-of-deep-sea-carbonate} \\
BLV3 & did you omit any other findings that others are reliant upon for their intrepetation? & 4 & Unfamiliar & ChatGPT & Sequestration and subduction of deep-sea carbonate in the global ocean since the Early Cretaceous & \url{https://pubs.geoscienceworld.org/gsa/geology/article-abstract/47/1/91/567642/Sequestration-and-subduction-of-deep-sea-carbonate} \\
BLV3 & use the visuals in the paper and make alt text for them, make sure you link the visual with the alt text so that a sighted person can verify the alt text, don't embed the alt text in the image, but produce it seperately & 2 & Unfamiliar & ChatGPT & Sequestration and subduction of deep-sea carbonate in the global ocean since the Early Cretaceous & \url{https://pubs.geoscienceworld.org/gsa/geology/article-abstract/47/1/91/567642/Sequestration-and-subduction-of-deep-sea-carbonate} \\
BLV3 & Figure 3 shows global visuals, add alt text to figure 3, make sure you include all information such as trend patterns, ledgers, axis units and scale, axis names, and title of the images, this is a global visual so use reference points to the continents that can be identified in the visual when you create the alt text, do not make up interpretations and do not add hallucinates. & 2 & Unfamiliar & Gemini & Sequestration and subduction of deep-sea carbonate in the global ocean since the Early Cretaceous & \url{https://pubs.geoscienceworld.org/gsa/geology/article-abstract/47/1/91/567642/Sequestration-and-subduction-of-deep-sea-carbonate} \\
BLV3 & Identify and name the subduction zones, pure carbonate platforms, and paleo-coastlines in figure 3 and describe where they are at in comparison to the results. Do not make up coastlines, paleo-coastlines, pure carbonate platforms, and subduction zones. Parameterize the information to fit with figure 3. Turnicate jargon when possible, don’t hallucinate. & 2 & Unfamiliar & Gemini & Sequestration and subduction of deep-sea carbonate in the global ocean since the Early Cretaceous & \url{https://pubs.geoscienceworld.org/gsa/geology/article-abstract/47/1/91/567642/Sequestration-and-subduction-of-deep-sea-carbonate} \\
BLV3 & Describe key finds in this pdf by Guy Munhoven, don’t hallucinate or make up results, don’t use information outside this paper to describe key finds. Add alt text to any visuals in the paper, make sure to include all details of these visuals such as global trends using geological names for features such as subduction zones, ridges, continents, shorelines, ocean basin, title all axises, give axis units, include graph titles, and ledger information, don’t make up figures or visuals & 1 & Familiar & Gemini & Model of Early Diagenesis in the Upper Sediment with Adaptable complexity – MEDUSA (v. 2): a time-dependent biogeochemical sediment module for Earth system models, process analysis and teaching & \url{https://www.researchgate.net/publication/352412122_Model_of_Early_Diagenesis_in_the_Upper_Sediment_with_Adaptable_complexity_-_MEDUSA_v_2_a_time-dependent_biogeochemical_sediment_module_for_Earth_system_models_process_analysis_and_teaching} \\
S4 & hey gemini, tell me about this paper - focus on what are the findings and what experiment paradigm and analysis were used. & 1 & Unfamiliar & Gemini & Low-dimensional criticality embedded in high-dimensional awake brain dynamics & \url{https://pubmed.ncbi.nlm.nih.gov/38669340/} \\
S4 & what is the key plot or figure that conveys the main finding of the paper & 2 & Unfamiliar & Gemini & Low-dimensional criticality embedded in high-dimensional awake brain dynamics & \url{https://pubmed.ncbi.nlm.nih.gov/38669340/} \\
S4 & could you summarize this paper to me focusing on the hypothesis first, methods used to test the hypothesis and key findings & 1 & Unfamiliar & ChatGPT & Low-dimensional criticality embedded in high-dimensional awake brain dynamics & \url{https://pubmed.ncbi.nlm.nih.gov/38669340/} \\
S4 & give me some intuitions for what avalanche analysis does and when it can be used. why is it suitable to apply here? & 3 & Unfamiliar & ChatGPT & Low-dimensional criticality embedded in high-dimensional awake brain dynamics & \url{https://pubmed.ncbi.nlm.nih.gov/38669340/} \\
S4 & what are the main plots that convey the main findings of this paper? & 2 & Unfamiliar & ChatGPT & Low-dimensional criticality embedded in high-dimensional awake brain dynamics & \url{https://pubmed.ncbi.nlm.nih.gov/38669340/} \\
S4 & help me understand figure 1J. what is PL ? & 2 & Unfamiliar & ChatGPT & Low-dimensional criticality embedded in high-dimensional awake brain dynamics & \url{https://pubmed.ncbi.nlm.nih.gov/38669340/} \\
S4 & could you interpret Fig 1J for me? & 2 & Unfamiliar & ChatGPT & Low-dimensional criticality embedded in high-dimensional awake brain dynamics & \url{https://pubmed.ncbi.nlm.nih.gov/38669340/} \\
S4 & in this paper, when they did factor analysis, did they report the dimensionality trends of readout population with learning? & 3 & Familiar & ChatGPT & Emergence of Coordinated Neural Dynamics Underlies Neuroprosthetic Learning and Skillful Control & \url{https://pubmed.ncbi.nlm.nih.gov/28190641/} \\
S4 & how is shared dimensionality different from overall dimensionality of readouts? & 3 & Familiar & ChatGPT & Emergence of Coordinated Neural Dynamics Underlies Neuroprosthetic Learning and Skillful Control & \url{https://pubmed.ncbi.nlm.nih.gov/28190641/} \\
S4 & yes please & 4 & Familiar & ChatGPT & Emergence of Coordinated Neural Dynamics Underlies Neuroprosthetic Learning and Skillful Control & \url{https://pubmed.ncbi.nlm.nih.gov/28190641/} \\
S4 & Could you point me to the key plots in the paper that convey the main findings? & 2 & Familiar & ChatGPT & Emergence of Coordinated Neural Dynamics Underlies Neuroprosthetic Learning and Skillful Control & \url{https://pubmed.ncbi.nlm.nih.gov/28190641/} \\
S4 & hey gemini, could you explain how the private and shared variance analysis could help us understand neural learning? & 3 & Familiar & Gemini & Emergence of Coordinated Neural Dynamics Underlies Neuroprosthetic Learning and Skillful Control & \url{https://pubmed.ncbi.nlm.nih.gov/28190641/} \\
S4 & did they look at overall dimensionality trends of readouts with learning? & 3 & Familiar & Gemini & Emergence of Coordinated Neural Dynamics Underlies Neuroprosthetic Learning and Skillful Control & \url{https://pubmed.ncbi.nlm.nih.gov/28190641/} \\
S4 & in neuroscience, overall dimensionality would be the degree of freedom that neural activity explores not the number of neurons themselves. With this context, could you tell me if the paper looked at overall readout dimensionality trends. How does Factor analysis separate variance ? & 3 & Familiar & Gemini & Emergence of Coordinated Neural Dynamics Underlies Neuroprosthetic Learning and Skillful Control & \url{https://pubmed.ncbi.nlm.nih.gov/28190641/} \\
S4 & what are the key plots that convey the main findings of the paper? help me understand them & 2 & Familiar & Gemini & Emergence of Coordinated Neural Dynamics Underlies Neuroprosthetic Learning and Skillful Control & \url{https://pubmed.ncbi.nlm.nih.gov/28190641/} \\
BLV4 & Can you give me a concise overview of this paper, especially methods and key findings? & 1 & Unfamiliar & Gemini & Somaesthetic Meditation Wearable: Exploring the Effect of Targeted Warmth Technology on Meditators’ Experiences & \url{https://dl.acm.org/doi/10.1145/3613904.3642557} \\
BLV4 & Can you elaborate on the study design a bit more? Specifically, what do you mean by "minimally-guided meditation sessions"? & 3 & Unfamiliar & Gemini & Somaesthetic Meditation Wearable: Exploring the Effect of Targeted Warmth Technology on Meditators’ Experiences & \url{https://dl.acm.org/doi/10.1145/3613904.3642557} \\
BLV4 & Please describe the pilot study from start to finish. In particular, I would like to learn more about how the research team determined possible body placement(s) for the device. & 3 & Unfamiliar & ChatGPT & Somaesthetic Meditation Wearable: Exploring the Effect of Targeted Warmth Technology on Meditators’ Experiences & \url{https://dl.acm.org/doi/10.1145/3613904.3642557} \\
BLV4 & I would now like a timeline-style summary of the main study itself. And if applicable, I would like you to note where each stage of the pilot studies was instrumental in crafting and refining the main study protocol. & 3 & Unfamiliar & ChatGPT & Somaesthetic Meditation Wearable: Exploring the Effect of Targeted Warmth Technology on Meditators’ Experiences & \url{https://dl.acm.org/doi/10.1145/3613904.3642557} \\
BLV4 & Can you please describe in detail the figures of this paper, what they aimed to show the reader, and why they were helpful in illustrating the authors' findings? & 2 & Familiar & Gemini & One vs. Many: Comprehending Accurate Information from Multiple Erroneous and Inconsistent AI Generations & \url{https://dl.acm.org/doi/10.1145/3630106.3662681} \\
BLV4 & No, I want to know what specific plots were used for each of these figures & 2 & Familiar & Gemini & One vs. Many: Comprehending Accurate Information from Multiple Erroneous and Inconsistent AI Generations & \url{https://dl.acm.org/doi/10.1145/3630106.3662681} \\
BLV4 & In Figure 3, each of the two plots have short, horizontal lines below each of the vertical lines for each condition. They also appear in different colors. What do these short, horizontal lines mean? Are they significant to understanding these plots and this figure? & 2 & Familiar & ChatGPT & One vs. Many: Comprehending Accurate Information from Multiple Erroneous and Inconsistent AI Generations & \url{https://dl.acm.org/doi/10.1145/3630106.3662681} \\
BLV4 & Why did the authors decide to use the means of Likert scale data in their visualizations? Do you think that this was the strongest way to visualize this data, or could there have been other measures that more appropriately captured insights from the Likert scale data? & 3 & Familiar & ChatGPT & One vs. Many: Comprehending Accurate Information from Multiple Erroneous and Inconsistent AI Generations & \url{https://dl.acm.org/doi/10.1145/3630106.3662681} \\
BLV5 & I am uploading a paper in PDF format that I got from ArXiV. Tell me how this paper relates to my work in aligning multimodal embeddings with neural activity. & 3 & Familiar & ChatGPT & Disentangling the Factors of Convergence between Brains and Computer Vision Models & \url{https://arxiv.org/abs/2508.18226} \\
BLV5 & I am not very familiar with DIONv3 models. Why do the authors used those specifically and how would that be compared with my current ViT\_l\_16 backbones. & 3 & Familiar & ChatGPT & Disentangling the Factors of Convergence between Brains and Computer Vision Models & \url{https://arxiv.org/abs/2508.18226} \\
BLV5 & Based on the attached pdf. Explain to me how it relates with my current research on brain alignment of multimodal models. & 3 & Familiar & Gemini & Disentangling the Factors of Convergence between Brains and Computer Vision Models & \url{https://arxiv.org/abs/2508.18226} \\
BLV5 & Describe for me this image from the paper: & 2 & Familiar & Gemini & Disentangling the Factors of Convergence between Brains and Computer Vision Models & \url{https://arxiv.org/abs/2508.18226} \\
BLV5 & I am uploading a 2021 neurIPS paper. I want you to: 1. extract the full text form the PDF. 2. carefully parce through the. 3. genrate a blind-friendly explanation on how this paper relates with my work. & 1 & Unfamiliar & Gemini & Attention Bottlenecks for Multimodal Fusion. & \url{https://arxiv.org/abs/2107.00135} \\
BLV5 & this is the file now & 4 & Unfamiliar & Gemini & Attention Bottlenecks for Multimodal Fusion. & \url{https://arxiv.org/abs/2107.00135} \\
BLV5 & generate a blind friendly, math and code focused of the figures of the paper. Explain to me how they were genrated, how would I present them on a lab meeting and how I could generate similar visuals & 2 & Unfamiliar & Gemini & Attention Bottlenecks for Multimodal Fusion. & \url{https://arxiv.org/abs/2107.00135} \\
S5 & Can you please summarize the different evaluation metrics of this benchmark? & 3 & Familiar & Gemini & BLADE: Benchmarking Language Model Agents for Data-Driven Science & \url{https://arxiv.org/abs/2408.09667} \\
S5 & Is the model querying different when measuring with coverage or average precision? & 3 & Familiar & Gemini & BLADE: Benchmarking Language Model Agents for Data-Driven Science & \url{https://arxiv.org/abs/2408.09667} \\
S5 & Is there any explicit suggestions of wanting coverage between runs? And what is the temperature? & 3 & Familiar & Gemini & BLADE: Benchmarking Language Model Agents for Data-Driven Science & \url{https://arxiv.org/abs/2408.09667} \\
S5 & What was the prompt for the agent fro the multiple choice example? & 3 & Familiar & Gemini & BLADE: Benchmarking Language Model Agents for Data-Driven Science & \url{https://arxiv.org/abs/2408.09667} \\
S5 & What is the text in Figure 19? & 2 & Familiar & Gemini & BLADE: Benchmarking Language Model Agents for Data-Driven Science & \url{https://arxiv.org/abs/2408.09667} \\
S5 & How about the prompt for analysis generation? & 3 & Familiar & Gemini & BLADE: Benchmarking Language Model Agents for Data-Driven Science & \url{https://arxiv.org/abs/2408.09667} \\
S5 & What are the prompts used for the agentic analysis generation? & 3 & Familiar & ChatGPT & BLADE: Benchmarking Language Model Agents for Data-Driven Science & \url{https://arxiv.org/abs/2408.09667} \\
S5 & What are examples of the different decision points and possible analysis decisions? & 3 & Familiar & ChatGPT & BLADE: Benchmarking Language Model Agents for Data-Driven Science & \url{https://arxiv.org/abs/2408.09667} \\
S5 & In the appendix, it talks about column pointer nodes where do those come from in the data? & 3 & Familiar & ChatGPT & BLADE: Benchmarking Language Model Agents for Data-Driven Science & \url{https://arxiv.org/abs/2408.09667} \\
S5 & In figure 4, what kinds of models do better and what do worse? & 2 & Familiar & ChatGPT & BLADE: Benchmarking Language Model Agents for Data-Driven Science & \url{https://arxiv.org/abs/2408.09667} \\
S5 & What decision points have the biggest precision range for model capabilities? & 2 & Familiar & ChatGPT & BLADE: Benchmarking Language Model Agents for Data-Driven Science & \url{https://arxiv.org/abs/2408.09667} \\
S5 & Who validates the validators? & 1 & Unfamiliar & Gemini & Who Validates the Validators? Aligning LLM-Assisted Evaluation of LLM Outputs with Human Preferences & \url{https://arxiv.org/abs/2404.12272} \\
S5 & What were some participant quotes related to criteria drift? & 3 & Unfamiliar & Gemini & Who Validates the Validators? Aligning LLM-Assisted Evaluation of LLM Outputs with Human Preferences & \url{https://arxiv.org/abs/2404.12272} \\
S5 & But what does it mean if criteria are always updating? How do they argue we can formulate correctness or evaluations? & 3 & Unfamiliar & Gemini & Who Validates the Validators? Aligning LLM-Assisted Evaluation of LLM Outputs with Human Preferences & \url{https://arxiv.org/abs/2404.12272} \\
S5 & What related work do they use to justify or support this criteria drift finding? & 3 & Unfamiliar & Gemini & Who Validates the Validators? Aligning LLM-Assisted Evaluation of LLM Outputs with Human Preferences & \url{https://arxiv.org/abs/2404.12272} \\
S5 & How does this differ from generic human in the loop model updating? & 3 & Unfamiliar & Gemini & Who Validates the Validators? Aligning LLM-Assisted Evaluation of LLM Outputs with Human Preferences & \url{https://arxiv.org/abs/2404.12272} \\
S5 & Is there anything discussed in the text that isn't shown in figure 1 (or vice versa) & 2 & Unfamiliar & ChatGPT & Who Validates the Validators? Aligning LLM-Assisted Evaluation of LLM Outputs with Human Preferences & \url{https://arxiv.org/abs/2404.12272} \\
S5 & Hmmm, instead of conceptual differences, I mean for the specific textual explanation of the EVALGEN Evaluation pipeline method, are there specific difference between the methods as described in the text and the Figure 1b diagram & 2 & Unfamiliar & ChatGPT & Who Validates the Validators? Aligning LLM-Assisted Evaluation of LLM Outputs with Human Preferences & \url{https://arxiv.org/abs/2404.12272} \\
S5 & In figure 2 part d, which part of figure 1 does it relate to? & 2 & Unfamiliar & ChatGPT & Who Validates the Validators? Aligning LLM-Assisted Evaluation of LLM Outputs with Human Preferences & \url{https://arxiv.org/abs/2404.12272} \\
S5 & In Table 1, Between the medical and product pipeline, what are important takeaways? & 2 & Unfamiliar & ChatGPT & Who Validates the Validators? Aligning LLM-Assisted Evaluation of LLM Outputs with Human Preferences & \url{https://arxiv.org/abs/2404.12272} \\
S5 & What are the downsides of this approach? & 3 & Unfamiliar & ChatGPT & Who Validates the Validators? Aligning LLM-Assisted Evaluation of LLM Outputs with Human Preferences & \url{https://arxiv.org/abs/2404.12272} \\
\end{longtable}
%TC:endignore

\end{document}